\documentclass[a4paper,12pt]{article}
\pdfoutput=1

\usepackage{jheppub}
\usepackage{bm,latexsym,amsmath,amssymb,amsfonts,mathtools,mathrsfs}

\usepackage{graphicx}
\usepackage{wrapfig}

\usepackage[hang]{subfigure}
\usepackage{color}

\definecolor{nicered}{rgb}{0.7,0.1,0.1}
\definecolor{nicegreen}{rgb}{0.1,0.5,0.1}

\usepackage{todonotes}
\definecolor{palatinate}{RGB}{128, 49, 123}

\newcommand{\be}{\begin{equation}}
\newcommand{\ee}{\end{equation}}
\newcommand{\beal}{\begin{aligned}}
\newcommand{\eeal}{\end{aligned}}
\newcommand\bea{\begin{eqnarray}}
\newcommand\eea{\end{eqnarray}}
\newcommand{\bec}{\begin{cases}}
\newcommand{\eec}{\end{cases}}

\DeclarePairedDelimiter\floor{\lfloor}{\rfloor}

\title{Thermodynamics of Many Black Holes}

\author[a,b,c]{Ruth Gregory}
\author[b]{Zheng Liang Lim}
\author[b]{Andrew Scoins}

\affiliation[a]{Department of Physics, King's College London,
The Strand, London WC2R 2LS, UK}
\affiliation[b]{Centre for Particle Theory, Durham University,
South Road, Durham, DH1 3LE, UK}
\affiliation[c]{Perimeter Institute, 31 Caroline Street North, Waterloo, 
ON, N2L 2Y5, Canada}
\emailAdd{ruth.gregory@kcl.ac.uk}
\emailAdd{andrew.d.scoins@durham.ac.uk}

\abstract{
We discuss the thermodynamics of an array of collinear black holes
which may be accelerating.
We prove a general First Law, including variations in the tensions
of strings linking and accelerating the black holes.
We analyse the implications of the First Law in a number of instructive cases,
including that of the C-metric,
and relate our findings to the previously obtained thermodynamics
of slowly accelerating black holes in anti-de Sitter spacetime.
The concept of thermodynamic length is found to be robust
and a Christoudoulou-Ruffini formula for the C-metric is shown.
}

\begin{document}
\maketitle

\section{Overview}\label{intro}

Black hole thermodynamics is a rich subject,
straddling both the classical and quantum aspects of gravity.
The thermodynamic charges of a black hole such as entropy and temperature,
while intrinsically quantum in nature,
are related to classical attributes such as horizon area and surface gravity 
\cite{Bekenstein:1973ur,Bekenstein:1974ax,Hawking:1974sw,Gibbons:1976}.
Indeed, it was considering the classical response of a black hole to 
infalling matter that led Bardeen, Carter, and Hawking to make the 
link between black hole variations and the First Law of thermodynamics in
their seminal paper \cite{Bardeen:1973gs}.

More recently, our understanding of black hole thermodynamics and
the interpretation of the various parameters has also been improving.
The first law of thermodynamics in gravitational systems
has been more comprehensively understood as an extended thermodynamical law
by including pressure in the guise of variations in vacuum 
energy \cite{Teitelboim:1985dp,Kastor:2009wy,Dolan:2010ha,Dolan:2011xt,
Kubiznak:2012wp}, and a more complete
understanding of the nature of ``$M$'' for the black hole
has emerged as the enthalpy of the system \cite{Kastor:2009wy}; see
\cite{Kubiznak:2016qmn} for a review.

These attempts at understanding the First Law
have largely considered single, isolated, black holes,
as in the Kerr-Newman family of solutions.
However, there are more complex, and therefore more interesting,
multi-black hole systems for which exact solutions are known.
Such geometries are thus amenable to thermodynamic analysis.
For example, the Israel-Khan solution \cite{IsraelKhan}
is an asymptotically flat geometry consisting 
of two black holes kept apart by a ``strut''---a
conical defect with an angular excess---corresponding
to a negative tension cosmic string.
More generally, one can sacrifice global asymptotic flatness
to remove the unphysical negative-tension defect
by running a positive tension cosmic string through the spacetime
\cite{AFV,AGK,GKW}.
In doing so, one retains local asymptotic flatness away from the core.
Generalising further, the \textit{accelerating black hole}, encoded in the 
C-metric \cite{Kinnersley:1970zw,Plebanski:1976gy}, consists of a black hole 
with a protruding cosmic string \cite{Gregory:1995hd}
(or an imbalance between antipodal strings) that provides an accelerating force. 
In this case, not only is asymptotic flatness lost near the string extending
to spatial infinity, but a non-compact \textit{acceleration horizon} forms.
Such systems beg the question:
how does one define thermodynamics for a geometry 
which is neither asymptotically flat, an isolated black hole,
nor (in the case of the Israel-Khan solution) stable?

Early thermodynamic investigations of black holes with conical defects
focused on a fixed deficit threading the horizon
\cite{AFV,Martinez:1990sd,Costa:2000kf,Dutta:2005iy,Herdeiro:2009vd},
or a deficit ``variation'' during the capture of a cosmic string \cite{Bonjour:1998rf}. 
The thermodynamic consequences of a truly varying deficit, however,
were not worked out until recently.
In particular, an accelerating, asymptotically locally anti-de Sitter
black hole has provided a context within which one maintains
excellent computational control.
This is owing both to one's ability to accelerate a black hole
without forming an acceleration horizon,
and the availability of the holographic dictionary \cite{Skenderis:2005}.
A fully general First Law was hence derived
\cite{Appels:2016uha,Appels:2017xoe},
accounting for a variation in a string's tension $\mu$:
\be
\delta M = T\delta S -\lambda\delta\mu + \cdots \,.
\ee
This tension comes paired with a conjugate thermodynamic potential $\lambda$,
christened the {\it thermodynamic length} of the string
\cite{Appels:2017xoe}.
These results were later generalised to accelerating black holes 
carrying rotational and $U(1)$ gauge charge
\cite{Anabalon:2018ydc,Anabalon:2018qfv}.
Interestingly, the expression for thermodynamic tension parallels that of 
the gravitational tension of Kaluza-Klein black strings 
\cite{Traschen:2001pb,Harmark:2004ch,Kastor:2012dt},
a set-up with no conical deficits.

Some understanding of the origin of thermodynamic length has also arisen.
Considering a system of two black holes coupled by a strut,
Krtou\v{s} and Zelnikov \cite{Krtous:2019} have found
a thermodynamic length corresponding to
the strut worldvolume evaluated at some fixed time.
This has since been verified for similarly coupled Kerr-Newman black holes
\cite{Ramirez-Valdez:2020,Garcia-Compean:2020}.

One should expect that if gravitational solutions
are truly representatives of a first law of thermodynamics
in the classical limit,
then one will find common features
no matter the number of black holes involved.
We demonstrate this here, by calculating variations of an array of 
collinear black holes -- connected by strings -- which
may be accelerated by external strings 
so as to form an acceleration horizon.
We allow all parameters in the solution to vary
and thereby prove a general First Law,
\be
\delta M = \sum_I T_I \delta S_I - \sum_J \lambda_J \delta\mu_J \,,
\label{eq:GFL}
\ee
wherein the temperatures $T_I$ and entropies $S_I$ of the 
compact black hole horizons contribute
together with the thermodynamic lengths $\lambda_J$ and tensions $\mu_J$
of the strings.
We justify the quantities appearing in \eqref{eq:GFL},
and consider its implications in a number of instructive cases,
including a triple black hole system and the C-metric geometry.
A key feature of our result is that the system behaves as a composite;
the individual black holes are not thermodynamically isolated, but 
each interacts with the other, a variation of one having implications 
for all the rest.

Note also that the First Law \eqref{eq:GFL} further supports the notion
of $M$ as enthalpy \cite{Kastor:2009wy}, even though there is no 
cosmological constant present here. The energy momentum of the
conical deficit, or cosmic string, takes the form of a worldsheet
cosmological constant: the string has a tension equal in magnitude and
opposite in sign to its energy density. Thus, the ``$-\delta\mu_J$'' term
in \eqref{eq:GFL} is in fact a ``$+\delta p_J$'' term, or pressure term, 
for the cosmic string. That the First Law contains a $\lambda \delta p$,
rather than $p\delta \lambda$ is indicative that $M$ truly represents an
enthalpy, and not an internal energy as previously imagined.

The outline of the paper is as follows:
In section \ref{sec:weyl}, we review the construction of black hole arrays
and acceleration horizons in Weyl gauge \cite{Weyl}.
In section \ref{sec:Array} we formulate a First Law 
for such systems, justifying the charges and potentials involved.
Section \ref{sec:multibhegs} discusses implications of the result
via some instructive examples and contains a novel Christodoulou-Ruffini 
mass formula \cite{Christodoulou:1972kt} for the C-Metric.

\section{Four dimensional Weyl metrics: black hole arrays}
\label{sec:weyl}

In this section we briefly review the multi-black hole solutions we
will be analysing. We will largely follow the presentation of \cite{Fay},
with minor notational changes. The main new result in this section is
a discussion of the determination of the acceleration scale for an array
of accelerating black holes in \eqref{eq:elllamdef}.
The black holes are aligned along an axis, and 
are static in the sense of possessing a time-like Killing isometry
in the region between the black hole and acceleration horizons.
Though an Israel-Khan-like solution for two rotating black holes is known
\cite{MankoRuiz:2019},
solutions for three or more Kerr black holes remain elusive.
We want to consider an arbitrary number of horizons
and thus will sidestep any discussion of rotation.
One expects that rotational charges may be included in the obvious way,
once an appropriate geometry is written down.

With temporal and axial symmetry, the metric can be written in a block
diagonal (Weyl) form,
with metric functions $\gamma,\nu,\text{ and }\alpha$
depending only on transverse coordinates $r$ and $z$:
\be
ds^2 = e^{2\gamma} dt^2 - e^{2(\nu-\gamma)} (dr^2 + dz^2)
- \alpha^2 e^{-2\gamma} d\phi^2 \,.
\label{4dweylmet}
\ee
The Einstein equations are:
\bea
\Delta \alpha &=& - 8\pi G \alpha e^{2(\nu-\gamma)} 
\left [ T^r_r + T^z_z \right ] \label{4dalph} \\
\Delta \gamma+ \frac{\nabla \gamma \cdot \nabla \alpha}{\alpha} &=&
4 \pi G e^{2(\nu-\gamma)} \left [
T^t_t - T^r_r - T^z_z - T^\phi_\phi \right ] \label{4dlam} \\
\Delta\nu + (\nabla\gamma)^2 &=& - 8\pi G e^{2(\nu-\gamma)} T^\phi_\phi
\label{4nu} \\
\frac{\partial_\pm^2\alpha}{\alpha} + 2 (\partial_\pm\gamma)^2
- 2 \partial_\pm \nu \frac{\partial_\pm\alpha}{\alpha} &=&
8\pi G [ T_{rr} - T_{zz} \pm 2iT_{rz} ] \label{4dc}
\eea
where $T_a^b$ is the energy momentum tensor of any bulk matter,
$\Delta$ is the two dimensional Laplacian $(\partial_r^2
+ \partial_z^2 = \partial_+\partial_-)$, with $\partial_\pm
= \partial_r \mp i\partial_z$ the derivatives with respect
to the complex coordinates $(r\pm iz)/2$.

In the absence of matter or a cosmological constant, these have a very elegant
solution: one simply fixes the conformal gauge freedom remaining in
the metric (\ref{4dweylmet}) by setting $\alpha \equiv r/K$, which
is consistent with (\ref{4dalph}). Note, we introduce the parameter $K$ 
here to maintain a $2\pi$ periodicity of the $\phi-$coordinate; 
this will become relevant when we discuss conical sources. 
With $\alpha\propto r$, (\ref{4dlam})
becomes a cylindrical Laplace equation for $\gamma$ in vacuo,
with solution
\be
\gamma = -2G \int
\frac{S({\bf r}') d^3 {\bf r}'}{ |{\bf r} - {\bf r}'|} 
\label{gammaintegral}
\ee
for a source with energy density $S({\bf r})$.
Note then that the metric component $\gamma$
is nothing but the Newtonian source of axial symmetry.
In turn, $\nu$ is determined from $\gamma$ via (\ref{4dc}).
Since the equation for $\gamma$ is linear, its solutions
can be superposed;
the nonlinearity of Einstein gravity shows up in the solution of $\nu$.
Note that, since regularity of the $r$-axis requires $\nu(0,z)=-\log K$,
in general there will be conical singularities when regular 
solutions for $\gamma$ are superposed.
These can be interpreted as strings or struts 
supporting the static sources in equilibrium. 

\subsection{The Schwarzschild solution}
\label{sec:schwarzschild}

As described in \cite{Fay},
a black hole may be represented by a finite-length line source\footnote{We 
make the gauge choice to centre the rod at $z=0$.},
$8\pi G S({\bf r}) = \delta(r)/r$ for $z\in[-m,m]$, yielding
\be
\gamma_S = -{1\over2} \int_{-m}^{m} {dz'\over[r^2 + (z-z')^2]^{1/2}}
= {\frac12} \log \frac{R_- - z_-}{R_+ - z_+} \,,
\label{Schlam}
\ee
where
\be
Z_\pm = z \mp m\qquad , \qquad R_\pm^2 = r^2 + Z_\pm^2 \,.
\ee
Integration of (\ref{4dc}) then gives
\be
\nu_S = \frac12 \log \frac{E_{+-}}{2R_+R_-} \,,
\label{Schnu}
\ee
where
\be
E_{+-} = R_+R_- + Z_+Z_- + r^2 \,.
\ee
Although this does not look like the familiar Schwarzschild black hole,
the simple transformation
\be
z=(\rho-m)\cos\theta \quad , \quad r^2 = \rho(\rho-2m) \sin^2\theta
\label{w2sch}
\ee
in fact returns the metric to its standard spherical form, with $2m=2GM_S$
representing the Schwarzschild radius.

\begin{wrapfigure}[16]{r}{0.26\textwidth}
\begin{center}
\vskip -22mm
\includegraphics[width=0.21\textwidth]{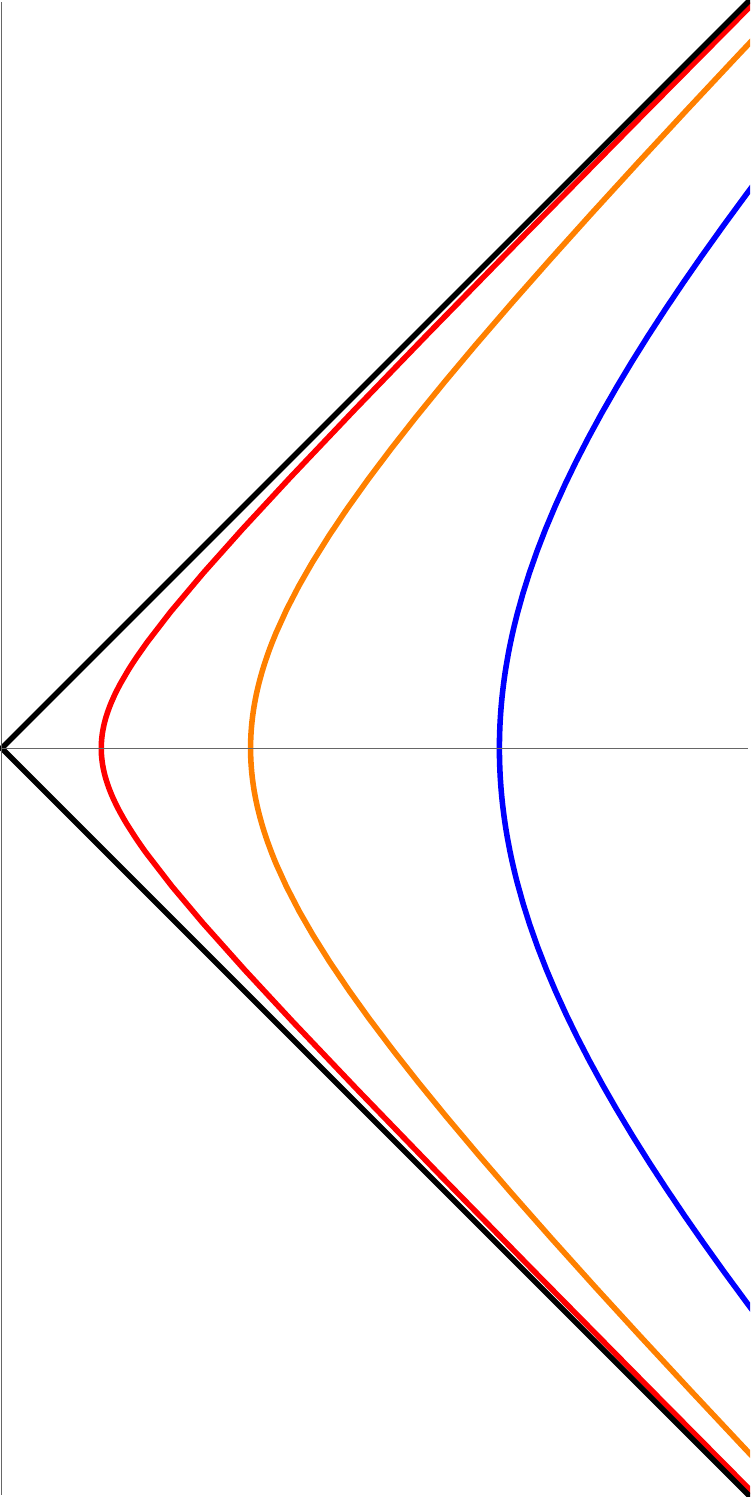}
\vspace{-10pt}
\end{center}
\caption{
Rindler worldlines of observers with differing accelerations
but same horizon.}
\label{fig:Rindler}
\end{wrapfigure}

\subsection{Rindler space}
\label{sec:Rindler}

Interestingly, one can formally introduce an
acceleration horizon by adding a semi-infinite line source (SILM)
\cite{EmparanReall:2002},
where $8\pi GS({\bf r}) = \delta(r)/r$ for $z>z_0$:
\be
\gamma_R = -{1\over2} \int_{z_0}^{\infty} {dz'\over[r^2 + (z-z')^2]^{1/2}}
\to  {\frac12} \ln \frac{R_0 - Z_0}{\ell_\gamma} \,,
\label{Rindlam}
\ee
where $Z_0 = (z-z_0)$, $R_0 = \sqrt{r^2 + Z_0^2}$, and the infinite integral
has been regulated by the lengthscale $\ell_\gamma$. Solving for $\nu$ 
yields the Rindler metric in Weyl coordinates:
\be
ds^2 =
\frac{(R_0-Z_0)}{\ell_\gamma} dt^2 - \frac{\ell_\gamma}{2R_0} [dr^2+dz^2]
- \frac{\ell_\gamma r^2}{(R_0-Z_0) } d\phi^2 \,.
\label{weylrindler}
\ee
Since Rindler spacetime is simply flat spacetime as observed by an 
accelerating observer, we can transform \eqref{Rindlam} to 
Minkowski spacetime (in cylindrical polars) via the transformation
\be
t = \frac{\ell_\gamma}{2} \log \left ( \frac{\tau+\zeta}{\zeta-\tau} \right)\quad,\quad
r = \frac{\rho}{\ell_\gamma} \sqrt{\zeta^2-\tau^2}\quad,\quad
z-z_0 = \frac{\tau^2+\rho^2-\zeta^2}{2\ell_\gamma}\, .
\ee
The origin of Minkowski corresponds to $z=z_0$, $r=0$,
(i.e.\ the start of the SILM), as expected.
The origin of the Weyl system
corresponds to $\zeta = \sqrt{2\ell_\gamma z_0}$,
which gives a natural choice of gauge for the Weyl system.
Note that the values of $z_0$ and
$\ell_\gamma$ are independent from the perspective of solving the Einstein
equations,
the former is a gauge choice---the origin of the $z$-coordinate---and
the latter because the same Rindler horizon can apply to observers with 
differing accelerations $A = 1/\ell_\gamma$; see figure \ref{fig:Rindler}.
Interpreting the origin of the Weyl system as
the location of the accelerating observer, thus
fixing the gauge, gives $z_0=1/2A$ from $\zeta = 1/A$.

\subsection{Many black holes}\label{sec:manybh}

\begin{figure}[h!] 
\centering
\includegraphics[width=\textwidth]{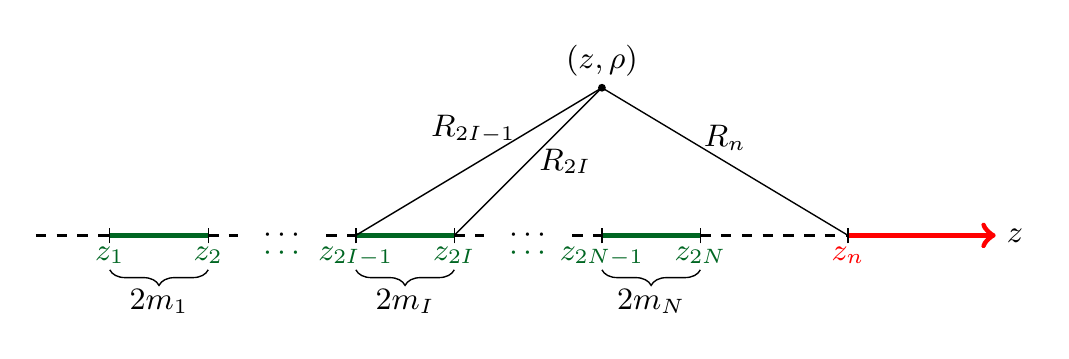}~
\caption{
The source arrangement for the multi-black hole system
of section \ref{sec:manybh}.
In the non-accelerating case, the point $z_n$,
representing the start of the SILM (thick red arrow),
and the SILM itself are absent;
its neighbouring string (dashed black)
instead extends to $z\rightarrow\infty$.}
\label{fig:AccSources}
\end{figure}

Now we can consider superposing solutions for $\gamma$, to
build up multi-black hole solutions as described in \cite{Fay}. 
We will briefly review these solutions, using a slightly different
notation to \cite{Fay} that is more suited to our argument.
Each black hole is represented by a rod of length $2m_I$, $I=1..N$,
and acceleration is represented by a SILM
as described above. We will label the rod ends 
at $z_i$, where $i=1..n$ and $z_1<z_2<..$. If we have an array of accelerating
black holes, $n=2N+1$, and the SILM begins at $z_n$, if we have an array of
(non-accelerating) black holes, then $n=2N$ is even.
This arrangement is depicted in figure \ref{fig:AccSources}

A natural generalisation of previous notation is
\be
\beal
Z_{i} &= z - z_i \,, &
R^2_{i} &= r^2 + Z_{i}^2 \,, \\
X_i &= R_i - Z_i \,,\qquad&
E_{ij} &= R_{i} R_{j} + Z_{i} Z_{j} +r^2\,.
\eeal
\label{rodends}
\ee
The solution for $\gamma$ is simply the superposition of 
the general potentials from \eqref{Schlam}, with $\nu$
then obtained by quadrature:
\be
\beal
\gamma &= \frac12 \sum_{i=1}^n (-1)^{i+1}
\log \frac{X_i}{\ell_\gamma} \,, \\
\nu &= \frac14 \sum_{i,j=1}^n (-1)^{i+j+1}  
\log \frac{E_{ij} }{\ell_\nu^2} + O_n \gamma \,.
\eeal
\label{multibhsol}
\ee
Here, the $\ell$'s are integration constants that cancel only if $n$ is even, 
and $O_n$ acts as a ``switch'' for  additional terms when $n$ is odd:
\be
O_n= n - 2 \floor*{ \frac{n}{2} } =
\begin{cases}
1 & n\;\text{odd} \\
0 & n\;\text{even} 
\end{cases} \,.
\ee

As we move to the thermodynamics of the system, we will need the limit
of these functions as we approach the axis, $r\to0$. We therefore conclude
this subsection by finding the behaviour of \eqref{multibhsol} as $r\to0$, and
discussing the conical deficits on the axis.
Noting that $R_i \to |Z_i|$ as $r\to0$, we see that
\be
X_i \sim |Z_i| - Z_i + \frac{r^2}{2|Z_i|}=
\begin{cases}
2|Z_i| & z < z_i \\
\frac{r^2}{2Z_i} & z > z_i
\end{cases} \,,
\label{Xlimit}
\ee
hence
\be
\gamma \sim \frac12 \sum_{i=1}^p (-1)^{i+1}
\log \frac{r^2}{2|Z_i| \ell_\gamma}
+ \sum_{i=p+1}^n (-1)^{i+1}
\log \frac{2|Z_i|}{\ell_\gamma} \;\;\;:\;\;\;\; z\in(z_p,z_{p+1})
\label{lamaxislimit}
\ee
where $p=0$ if $z< z_1$ leaving only the second sum, and conversely
the first sum for $z>z_n$.

Next,
\be
E_{ij} \sim \begin{cases}
2 Z_i Z_j & z< \text{Min}[z_i,z_j] \; , \;\; z> \text{Max}[z_i,z_j] \\
\frac{r^2(z_i-z_j)^2}{2|Z_iZ_j|} & \quad\text{Min}[z_i,z_j] < z < \text{Max}[z_i,z_j]
\end{cases} \,.
\label{Elimit}
\ee
Hence if we approach the axis at the $I\textsuperscript{th}$ black hole, 
for which $z\in(z_{2I-1},z_{2I})$,
\be
\beal
\nu \sim & 
\frac14 \sum_{i,j=1}^{2I-1} (-1)^{i+j+1} \log\left ( \frac{2Z_iZ_j}{\ell_\nu^2}\right)
+ \frac14 \sum_{i,j=2I}^{n} (-1)^{i+j+1} \log\left ( \frac{2Z_iZ_j}{\ell_\nu^2}\right)\\
&
+ \frac12 \sum_{i=1}^{2I-1} \sum_{j=2I}^n (-1)^{i+j+1} 
\log\left ( \frac{r^2(z_i-z_j)^2}{2|Z_iZ_j| \ell_\nu^2}\right)
+ O_n\gamma  \,.
\eeal
\label{nubhaxislimit}
\ee
Away from the black holes, writing $\nu_0 = \frac12 \log\left ( 
\frac{\sqrt{2} \ell_\nu}{\ell_\gamma} \right)$, we have:
\be
\beal
\nu(0,z) &= O_n\nu_0
&z<z_1 \,\\
&=\sum_{i=1}^{2I} \sum_{j=2I+1}^n (-1)^{i+j+1}
\log(z_j-z_i)  + O_n \nu_0
& \quad z_{2I}<z<z_{2I+1} \, \\
&=O_n \left [ \sum_{i=1}^{n-1} (-1)^{i}
\log(z_n-z_i)  +\nu_0 \right]
& z_{2N} < z <z_n \,.
\eeal
\label{nuconaxislimit}
\ee
Notice that $z_j-z_{i+1} < z_j - z_i<z_{j+1}-z_i$, thus $\nu(0,z)<\nu_0$
for the first string tension, and (for accelerating black holes) $\nu_N<\nu_0$.

We can now identify the conical structure on the axis.
The axis will have a conical defect if the circumference of circles 
of proper radius $\Delta r$ around it are not $2\pi \Delta r$. For small
$r$, $\Delta r \sim e^{\nu(0,z)-\gamma(0,z)}$ and the circumference
is $2\pi r e^{-\gamma(0,z)} /K$, hence the deficit angle $\delta$ is 
\be
\delta = 2 \pi \lim_{r\to0} \left [ 1 - \frac{e^{-\nu(0,z)}}{K} \right] \,,
\ee
which is related to the cosmic string tension via $\delta = 8\pi G \mu$.
\eqref{nuconaxislimit} dictates how the deficit angle changes as we
move between the black holes.
The tension between the 
$I\textsuperscript{th}$ and $(I+1)\textsuperscript{th}$ black hole is
\be
\mu_I = \frac14 \left ( 1 - \frac{e^{-O_n\nu_0}}{K} \prod_{i=1}^{2I}
\prod_{j=2I+1}^n (z_j-z_i)^{(-1)^{i+j}}
\right) \,.
\label{muIdef}
\ee
The final black hole has $\mu_N$ as the deficit for $z>z_{2N}$, 
\be
\mu_N = 
\begin{cases}
\frac14 \left ( 1 - \frac{1}{K} \right ) & n=2N \\
\frac14 \left ( 1 - \frac{e^{-\nu_0}}{K} 
\prod_{i=1}^{N} \frac{(z_n-z_{2i-1})}{(z_n-z_{2i})}
\right ) & n = 2N+1
\end{cases} \;,
\label{muNdef}
\ee
and for the incident tension, $z<z_1$, we have
\be
\mu_0 = \frac14 \left ( 1 -  \frac{e^{-O_n\nu_0}}{K} \right) \,.
\label{mu0def}
\ee

We now see the interpretation of $K$. For the non-accelerating black hole
array, there is an ambient tension running through the system, as 
the deficit outside the array ($z<z_1$ and $z>z_{2N}$) have the same
conical deficit of 
\be
\mu_0 = \mu_N = \frac14 \left ( 1 -  \frac1{K} \right) \,.
\ee
Equation \eqref{nuconaxislimit} shows that $e^{\nu(0,z)}<1$
between the black holes.
Hence, if we did not insert the parameter $K$,
instead retaining a $2\pi$ periodicity of $\phi$ 
for $z<z_1$ and $z>z_{2N}$,
the conical singularity between any two of the black holes 
would be an excess $\delta<0$,
corresponding to a negative tension ``cosmic strut''
as in \cite{Krtous:2019}.
Although one can consider such systems
\cite{Costa:2000kf,Krtous:2019,Ramirez-Valdez:2020,Garcia-Compean:2020},
we prefer to keep physical sources.
We therefore take $K$ large enough that all 
the conical singularities are deficits
and correspond, in principle, to physical cosmic strings 
\cite{AGK,GKW}.
Note however that if $K>1$,
there is an ambient conical deficit through the spacetime,
irrespective of whether there is acceleration.

For an accelerating black hole array, we follow the convention of
\cite{Anabalon:2018ydc,RuthAndy}
that $K$ measures the ambient deficit, i.e.
\be
\mu_0 + \mu_N = \frac12 \left ( 1 - \frac1K \right) \,.
\ee
This in turn allows us to determine $\nu_0$:
\be 
e^{\nu_0}= \left ( \frac{\sqrt{2} \ell_\nu}{\ell_\gamma} \right)^{1/2}
= \frac12 \left ( 1+ \prod_{i=1}^{N} \frac{(z_n-z_{2i-1})}{(z_n-z_{2i})} \right)
\equiv \frac12 \left ( 1+ V_n \right) \,.
\label{nu0def}
\ee
thus we have
\be
\mu_0 =  \frac14 \left ( 1 - \frac2{(1+V_n) K} \right) \quad,\qquad
\mu_N =  \frac12 \left ( 1 - \frac{2V_n}{(1+V_n)K} \right) \,.
\label{themus}
\ee

Note however that the choice of $K$ is not unique; this one,  \eqref{themus},
corresponds to the same normalisation as the standard C-metric, however, 
if one were viewing the metric as a split cosmic string, then an alternate 
natural choice might be to normalise the ``initial'' deficit.
That is, we could choose
$\mu_0 = \frac14 \left ( 1 - \frac1K \right)$,
in which case
$\mu_N = \frac14 \left ( 1 - \frac{V_n}K \right)$.

Finally, we are left with the length scale $\ell_\gamma$,
which is (only) present in an accelerating array,
This parameter represents the net acceleration scale of the spacetime.
We expect that for small accelerations (large $z_n$) 
this should asymptote the Rindler value $\ell_\gamma\sim 2z_n$. 
Interpreting the acceleration as
the overall mass of the composite black hole system divided by the overall
force measured by the differential deficit, we are led to
\be
\ell_\gamma = \frac{M}{\mu_0-\mu_N}
= \frac{V_n+1}{V_n-1} \sum_1^N (-1)^k z_k \;\;,
\label{eq:elllamdef}
\ee
where $M = \sum m_I/K$ is the total mass of the system 
(see section \ref{sec:thermoparams}).
We see that $\ell_\gamma$ has the required large $z_n$ limit
and a clear physical interpretation in close analogy 
with its pure Rindler cousin from section \ref{sec:Rindler}.

\section{Thermodynamics of an array of black holes}
\label{sec:Array}

We now derive a First Law for collinear black holes
with varying positive tension strings
and a possible acceleration horizon,
the solutions for which were presented in section \ref{sec:manybh}.

\subsection{Deriving the thermodynamic parameters}
\label{sec:thermoparams}

First we need to derive these relevant thermodynamic parameters.
For the entropy of a given black hole, we compute the area of the
relevant horizon
\be
S_I = \lim_{r\to0} \, 
\frac{\pi}{2K} \int_{z_{2I-1}}^{z_{2I}} r e^{\nu-2\gamma} dz
= \frac{\pi m_I}{K}\lim_{r\to0} \, r e^{\nu-2\gamma} \,.
\label{eq:entropy}
\ee
For the temperature, the standard techniques apply, yielding
\be
T_I =  \lim_{r\to0} \, \frac1{2\pi}\, \frac{e^{2\gamma}}{r e^\nu}
=\frac{m_I}{2KS_I} \,.
\label{eq:temperature}
\ee
The limit of $re^{\nu-2\gamma}$ as we approach the axis is given by
\eqref{lamaxislimit}, \eqref{nubhaxislimit}, and using \eqref{nu0def}, we obtain:
\be
\beal
\log (re^{\nu-2\gamma}) &\to
\log 2 + O_n \log\left(\frac{\ell_\gamma e^{\nu_0}}{2}\right) 
+ \sum_{i=1}^{2I-1} \sum_{j=2I}^n (-1)^{i+j+1} \log |z_j-z_i| \,.
\eeal
\ee

The most challenging thermodynamic quantity to identify is the total mass.
This is in part due to the fact that external strings which extend to infinity
prevent global asymptotic flatness and thus render the ADM mass
\cite{ADM:1959} ill-defined.
The presence of a non-compact acceleration horizon further complicates matters.
Some attempt has been made \cite{Dutta:2005iy}
to redefine ADM mass in the presence of a conical defect
by calculating the mass relative to conical Minkowski space,
rather than pure Minkowski as one would in the usual construction.
However, such a construction gives undesirable results.
In particular, one would conclude that the mass of the C-metric is vanishing.
This is puzzling from the perspective of having no smooth transition to 
the non-accelerating black hole.
It is also counter to the intuition gained from 
the slowly accelerating black hole in AdS, for which the mass 
(with an appropriately normalised time coordinate)
is $M\textsubscript{AdS}=m/K$.
One may be confident in the AdS calculation
due to the holographic correspondence.

Although one may struggle to find a useful notion of ADM mass,
the existence of the $\partial_t$ isometry means that one
still has a Komar construction \cite{Komar:1963} at one's disposal.
Focusing first on the non-accelerating case,
\cite{Costa:2000kf} calculated the ADM mass 
for a system of collinear black holes without external strings 
($\mu_0=\mu_N=0$).
One can compute the asymptotic behaviour,
\be
\beal
&e^{2\gamma} \sim 1-\frac{2(\Sigma_{I=1}^{N}m_I)}{\tilde r}
+\mathcal{O}(\tilde r^{-2}) \,,\qquad
&\nu \sim \mathcal{O}(\tilde r^{-2}) \,,
\eeal
\ee
where $\tilde{r}$ is a suitable radial coordinate,
and simply read off the mass.
As discussed above, when we have an ambient conical deficit
the ADM mass is undefined, but we may instead read off the Komar mass
as $M=\sum_{I=1}^{N}m_I/K$.

When an acceleration horizon is present,
the situation requires more explanation.
We take $k=\partial_t$ as our Killing vector field 
generating time translations.
The normalisation of $k$ is implicit in the choice \eqref{eq:elllamdef}
of $\ell_\gamma$; 
see the discussion given at the end of section \ref{sec:weyl}.
The covector associated to $k$ is $k^\flat=e^{2\gamma}dt$.
Taking the exterior derivative and Hodge dual, we find
\be
\star dk^\flat = \frac{r}{Ke^{2\gamma}}
\left[(\partial_r e^{2\gamma})dz -(\partial_z e^{2\gamma})dr \right]
\wedge d\phi \,.
\ee
The causal structure of the spacetime is now significantly more complicated 
than in the non-accelerating case,
but there is still a well defined spatial infinity \cite{PavelInterpreting}.
To calculate the total mass,
one could, in principle, integrate this form 
over a two-surface there.
That said, it is more instructive to use Gauss' law
to rewrite the boundary integral 
as the sum of integrals over each black hole horizon and a bulk integration:
\begin{equation}
\frac{1}{8\pi}\int_{\infty}\star dk^\flat =
\frac{1}{8\pi}\sum_{I=1}^{N}\int_{\mathcal{H}_I}\star dk^\flat
+ M\textsubscript{bulk} \,.
\label{integrand}
\end{equation}
The quantity on the left hand side is the total 
mass\footnote{There is a caveat here that we have divided through to 
retain only the mass of objects on one side of the acceleration horizon.} $M$.
From \eqref{lamaxislimit} and \eqref{nubhaxislimit},
we have the relevant behaviour for the integrand on the right hand side of 
\eqref{integrand} near the $I\textsuperscript{th}$ horizon $\mathcal{H}_I$,
\be\beal
&\partial_r e^{2\gamma}
\sim \frac{r}{2\left\vert z_{2I+1}z_{2I}\right\vert} \,, &\qquad
&e^{-2\gamma} \sim \frac{4\left\vert z_{2I+1}z_{2I}\right\vert}{r^2} \,,
\eeal\ee
making the integrand straightforward:
\be
\lim_{r\rightarrow 0}\left[\star dk^\flat\right]_{z\in(z_{2I},z_{2I+1})}
= \frac{2}{K}dz\wedge d\phi + \ldots \,.
\ee
Hence we conclude that the integral over $\mathcal{H}_I$,
which we interpret as the mass of an individual black hole in the array, is
\be
M_I \equiv \frac{1}{8\pi}\int_{\mathcal{H}_I}\star dk^\flat
= \frac{m_I}{K} \,.
\ee
Finally, we note that the volume integral $M\textsubscript{bulk}$ vanishes,
and that the strings themselves make no contribution to the above calculation.

The conclusion is that the total Komar mass is directly related 
to the rod lengths of compact horizons.
The same result for the mass of the solitary accelerating black hole 
has been proposed in \cite{Anabalon:2018qfv}, 
albeit with a non-commital attitude to the normalisation of $k$.
We also observe a clear similarity with the holographically calculated
mass of a slowly accelerating black hole in AdS \cite{Anabalon:2018ydc}.

\subsection{The First Law of Thermodynamics}\label{sec:firstlaw}

We now show how to derive equation \eqref{eq:GFL},
the first law of thermodynamics for an array of collinear black holes.
Consider a variation to the array.
The solution \eqref{multibhsol} describes a coupled system;
any variation of one black hole will impact on all the others.
Therefore, we do not expect individual First Laws for each black hole.
Instead, it makes sense to consider a variation of the total mass
\be
M = \sum_{I=1}^{N} \frac{m_I}{K} \,,
\ee
as this is a state function of the complete system.
Indeed, this is the philosophy for the First Law derived in \cite{Krtous:2019}.
Thus, to derive a First Law, we must compute
\be
\delta M = \sum_{I=1}^N \frac1K \delta m_I - m_I \frac{\delta K}{K^2} \,.
\ee

We begin by computing the variation in entropies
for the individual black holes:
\be
\sum_{I=1}^N T_I \delta S_I =
\frac12 \sum_{I=1}^N \delta \left(\frac{m_I}{K}\right) + S_\Sigma
+ \frac{ O_nM}{2}
\left ( \frac{\delta\ell_\gamma}{\ell_\gamma} + \delta\nu_0\right) \,,
\label{tds}
\ee
having replaced $\ell_\nu = e^{2\nu_0} \ell_\gamma/\sqrt{2}$, and where
\be
S_\Sigma =
\sum_{I=1}^N \frac{m_I}{2K} \sum_{i=1}^{2I-1} \sum_{j=2I}^{2N} 
(-1)^{i+j+1} \frac{\delta (z_j - z_i)}{z_j-z_i} 
+ O_n \sum_{I=1}^N \frac{m_I}{2K} \sum_{i=1}^{2I-1} 
(-1)^{i} \frac{\delta (z_n - z_i)}{z_n-z_i} \\ \,.
\label{ssigmadef}
\ee
This contains part of what we need for a First Law, but has a rather 
messy sum!

Now we turn to the cosmic strings. We write the thermodynamic 
lengths for the strings as $\lambda_I = - e^{\nu_I} L_I $, and then
vary the tensions in \eqref{muIdef}, \eqref{muNdef}, and \eqref{mu0def} 
to obtain the contribution to the First Law coming from the tensions:
\be
- \sum_{I=0}^{N} \lambda_I \delta \mu_I = \sum_{I=0}^{N}
\frac{L_I\delta K}{4K^2} + O_n \sum_{I=0}^{N}
\frac{L_I}{4K} \delta \nu_0 +\mu_\Sigma \,,
\ee
where
\be
\mu_\Sigma =
\sum_{I=1}^{N-1} \frac{L_I}{4K} \sum_{i=1}^{2I} \sum_{j=2I+1}^{2N} 
(-1)^{i+j+1} \frac{\delta (z_j - z_i)}{z_j-z_i} 
+ O_n \sum_{I=1}^{N} \frac{L_I}{4K} \sum_{i=1}^{2I} (-1)^i 
\frac{\delta(z_n-z_i)}{z_n-z_i} \,.
\label{musigmadef}
\ee

Putting these two expressions together, we have:
\be
\beal
\sum_{I=1}^N &T_I \delta S_I
- \sum_{I=0}^{N} \lambda_I \delta \mu_I  \\
&= \frac{\delta M}2 + S_\Sigma + \mu_\Sigma
+ \sum_{I=0}^{N} \frac{L_I\delta K}{4K^2}
+ \frac{ O_n}{4K} \sum_{I=0}^{N} (L_I+2m_I) \delta \nu_0 
+ 2m_I  \frac{\delta\ell_\gamma}{\ell_\gamma}  \,.
\label{prelimfirstlaw}
\eeal
\ee

First, let us deal with the sums $S_\Sigma$ and $\mu_\Sigma$
in these expressions. Observing that $m_I = (z_{2I}-z_{2I-1})/2$,
we can rewrite the entropy sum as 
\be
S_\Sigma =
\sum_{k=1}^{2N} \sum_{i=1}^{2[\frac{k+1}{2}] -1} 
\!\!\!\! \sum_{j=2[\frac{k+1}{2}]}^{2N}
\frac{(-1)^{i+j+k+1}}{4K} z_k \frac{\delta (z_j - z_i)}{z_j-z_i}
+ \frac{O_n}{4K}  \sum_{k=1}^{2N}\sum_{i=1}^{2\left [ \frac{k+1}{2} \right] -1} 
(-1)^{i+k} z_k \frac{\delta (z_n - z_i)}{z_n-z_i} 
\ee
Generalising \cite{Krtous:2019} for the thermodynamic lengths
of strings in between horizons as $L_I = z_{2I+1}-z_{2I}$
(with the exception of $L_0$ and $L_N$ -- see later) gives the tension sum as
\be
\beal
\mu_\Sigma &= \sum_{k=2}^{2N-1}\  \sum_{i=1}^{2[\frac{k}{2}] }
\sum_{j=2[\frac{k}{2}]+1}^{2N}
\frac{(-1)^{i+j+k}}{4K} z_k \frac{\delta (z_j - z_i)}{z_j-z_i}\\
&+ \frac{O_n}{4K} \sum_{k=1}^{2N-1}\
\sum_{i=1}^{2\left [ \frac{k}{2} \right] }  (-1)^{i+k+1} z_k
\frac{\delta(z_n-z_i)}{z_n-z_i}
+ \frac{L_N}{4K} \sum_{i=1}^{2N} (-1)^i \frac{\delta(z_n-z_i)}{z_n-z_i} \,.
\eeal
\ee
We now see that many of the terms in $S_\Sigma$ are cancelled by
terms in $\mu_\Sigma$, leaving just $k=1,2N$ from the entropy sum, 
and intermediate $i,j$ terms from each
when $2[\frac{k+1}{2}]$ differs from $2[\frac{k}{2}]+1$:
\be
\beal
&S_\Sigma+\mu_\Sigma \\
&= \sum_{j=2}^{2N} \frac{(-1)^{j+1}}{4K} z_1 \frac{\delta (z_j - z_1)}{z_j-z_1}
+\sum_{i=1}^{2N -1} \frac{(-1)^{i+1}}{4K} z_{2N} 
\frac{\delta (z_{2N} - z_i)}{z_{2N}-z_i} \\
&+ \sum_{k=2}^{2N-1}\left [ \sum_{j=k+1}^{2N}
\frac{(-1)^{j+k}}{4K} z_k \frac{\delta (z_j - z_k)}{z_j-z_k}
+ \sum_{i=1}^{k-1} \frac{(-1)^{i+k+1}}{4K}  z_k
\frac{\delta (z_i - z_k)}{z_i-z_k}\right]\\
&+ \frac{O_n}{4K}  \left ( \sum_{i=1}^{2N-1} 
\left [ (-1)^i (L_N+z_{2N}) -z_i \right ]\frac{\delta(z_n-z_i)}{z_n-z_i}
+ L_N \frac{\delta(z_n-z_{2N})}{z_n-z_{2N}} \right) \\ 
&= \sum_{j=2}^{2N} \sum_{i=1}^{j -1} \frac{(-1)^{i+j+1}}{4K}(\delta z_j - \delta z_i)
+ \frac{O_n}{4K} \sum_{k=1}^{2N} 
(-1)^k  (L_N+z_{2N}-z_k) \frac{\delta(z_n-z_k)}{z_n-z_k}  \,.
\eeal
\ee

We now have to identify $L_N$ (and $L_0$).
We write
\be
L_N = z_c - z_{2N} \qquad,\qquad L_0 = z_1 - z_c
\ee
in keeping with the expressions for $L_I$, where $z_c$ is a normalisation,
similar to that of the SILM in $\gamma$, to be determined.
We can therefore reduce this combination to
\be
\beal
S_\Sigma+\mu_\Sigma &= \sum_{I=1}^{N}\frac{\delta m_I}{2K}
+ \frac{O_n}{4K} \sum_{k=1}^{2N}  (-1)^k \left ( \delta(z_n-z_k) 
+  (z_c-z_n)  \frac{\delta(z_n-z_k)}{z_n-z_k}\right) \\
&= \sum_{I=1}^{N}\frac{\delta m_I}{2K} \left ( 1 - O_n \right)
+ \frac{O_n}{4K}  (z_c-z_n) \sum_{k=1}^{2N} 
(-1)^k  \frac{\delta(z_n-z_k)}{z_n-z_k}\\
&= \sum_{I=1}^{N}\frac{\delta m_I}{2K} \left ( 1 - O_n \right)
- \frac{O_n}{4K}  (z_c-z_n) \frac{\delta V_n}{V_n} \,.
\eeal
\ee

Having simplified $S_\Sigma+\mu_\Sigma$, we now turn to the rest of
the putative First Law, \eqref{prelimfirstlaw}. We note that the sum of the
thermodynamic lengths can be related to the sum of the masses:
\be
\sum_{I=0}^{N} L_I = \sum_{I=1}^{N-1} (z_{2I+1}-z_{2I}) +L_N +L_0
= - 2 \sum_{I=1}^{N} m_I  \,.
\ee
Hence,
\be
\beal
\sum_{I=1}^N T_I \delta S_I -\sum_{\mu's} \lambda_I \delta \mu_I 
&=  \frac{\delta M}2 + \sum_{I=0}^{N}
\frac{L_I\delta K}{4K^2}+ \sum_{I=1}^{N}\frac{\delta m_I}{2K}\\
& + \frac{O_n}{4K} \left [ \sum_{I=1}^N 2m_I  \frac{\delta\ell_\gamma}{\ell_\gamma}
- \sum_{I=1}^{N} 2\delta m_I - (z_c-z_n) \frac{\delta V_n}{V_n} \right] \\
&= \delta M+ \frac{O_n}{2K} \left [ m_{TOT}  \frac{\delta\ell_\gamma}{\ell_\gamma}
- \delta m_{TOT} - (z_c-z_n) \frac{\delta V_n}{2V_n} \right]  \,,
\eeal
\ee
where $m_{TOT} = \sum_I m_I$ is shorthand for the sum of the individual rod
lengthscales.
Thus, we have derived the First Law \eqref{eq:GFL}
for a general array of black holes,
provided we identify
\be
z_c = z_n + \frac{2m_{TOT}  \delta\ell_\gamma / \ell_\gamma
- 2\delta m_{TOT} }{\delta V_n/V_n}
= z_n - \frac{4V_n m_{TOT}}{V_n^2-1} \,,
\ee
for the accelerating black hole. For the non-accelerating black hole, the
First Law is automatically satisfied and we
set $L_0=L_N=(z_1+z_{2N})/2$.

\section{Exploring Multi-Black Hole Spacetimes}
\label{sec:multibhegs}

Having derived these expressions, it is interesting to explore 
some sample accelerating and non-accelerating black hole 
arrays to gain an understanding of the interdependency of 
black hole entropy, and to see how the strings contribute
to the thermodynamic system as well as cross-checking
against known results.

\subsection{Non-accelerating arrays}

We start by considering non-accelerating black holes.
This includes the Schwarzschild case as a basic cross-check of our results,
and the two black hole system which has already been considered
in the literature 
\cite{Costa:2000kf,Krtous:2019,Ramirez-Valdez:2020,Garcia-Compean:2020}.

\subsubsection{Schwarzschild with a string}

As discussed in section \ref{sec:weyl},
the Schwarzschild solution (with an axial conical defect)
has $n=2$, $N=1$, and $z_2-z_1=2m$.
Conventionally, we set the centre of the rod at the origin
so that $z_2=-z_1=m$.
From \eqref{eq:entropy} and \eqref{eq:temperature}
we find that the entropy and temperature are
$S=4\pi m^2/K$ and $T=1/8\pi m$ respectively, as expected.
For the cosmic string piercing the horizon, we have
$\mu_0=\mu_1=\frac14\left( 1 - \frac1K \right)$, and
$\lambda_0=\lambda_1=m$
in agreement with \cite{Appels:2017xoe}.

\subsubsection{Two black holes}

The First Law for the two black hole system, with $K=1$,
was explored in \cite{Krtous:2019}.
This value of $K$ means that there are no strings running to infinity, but instead
the black holes are held apart by a negative tension strut.
Nevertheless, for larger $K$, 
\be
K \geq \frac{D^2-(m_2-m_1)^2}{D^2- (m_1+m_2)^2}
\ee
where $D$ is the distance between the centres of the two rods,
we find results harmonious with their conclusions:
the First Law holds with the thermodynamic length of the defect 
connecting the black holes given by
the worldsheet volume of the string per unit time.
The thermodynamic lengths of the semi-infinite strings are now
\be
\lambda_0 = \lambda_2 = \frac{z_4-z_1}{2} = \frac{D}2 + \frac{m_1+m_2}2
\ee
i.e.\ the system responds to the average mass, and the length between
the black holes. Note that $\lambda_1 = - (z_3-z_2) e^{\nu_1}$ also has
a factor of the separation that is important for consistency in varying
the net conical deficit of the system. We discuss this in more detail below
for three black holes.

\subsubsection{Three black holes}

\begin{figure}[h] 
\centering
\includegraphics[width=\textwidth]{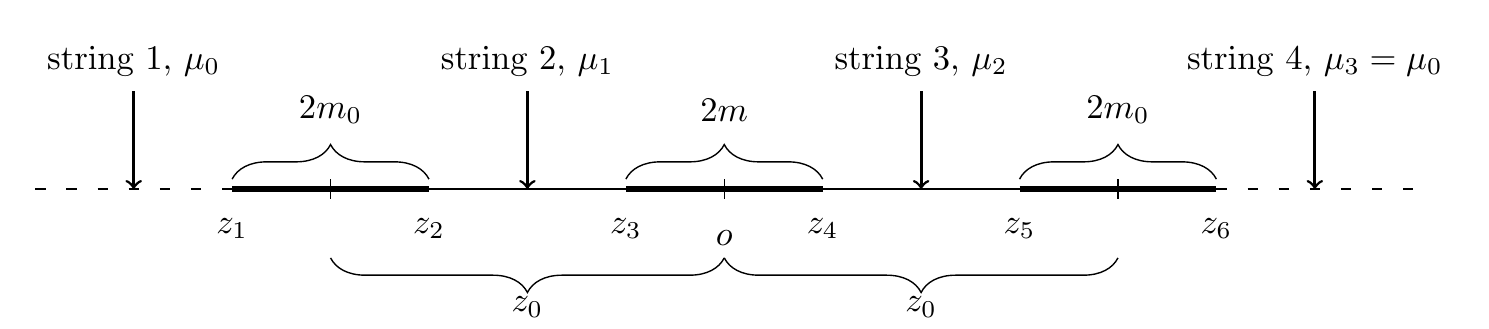}
\caption{The source arrangement for three (non-accelerating) black holes.}
\label{fig:threerods}
\end{figure}

The three black hole system has rods on the intervals
$(z_1,z_2)$, $(z_3,z_4)$, and $(z_5,z_6)$; see figure \ref{fig:threerods}. 
We are interested in exploring how the locations of the sources 
affect entropy and tension,
and how a perturbation of one black hole impacts on the others.
Hence, we consider a set-up in which 
the two outer black holes have equal mass
and spacing from the
middle black hole, which is centred around the origin:
$z_6-z_5=z_2-z_1=2m_0$, and $z_6=-z_1=z_0$, $z_4=-z_3=m$.
The entropies and tensions then become:
\be
\beal
S_1 &= \frac{4\pi m_0^2}{K}
\frac{(z_0+m_0)(z_0+m_0+m)}{z_0(z_0+m_0-m)} =S_3 \\
S_2 &= \frac{4\pi m^2}{K}
\frac{(z_0^2-m_0^2)(z_0+m_0+m)^2}{z_0^2(z_0-m_0+m)^2} \\
\mu_0 &= \frac14\left ( 1 - \frac1K \right) \\
\mu_1 &= \frac14\left ( 1 - \frac{z_0^2(z_0^2-(m_0-m)^2)}
{(z_0^2-m_0^2)(z_0^2-(m_0+m)^2)K} \right) =\mu_2  \,.
\eeal
\ee
It is easy to see that $\mu_1<\mu_0$.
This is to be expected:
in order to retain equilibrium,
additional force must be applied on the outer black holes to
counterbalance their attraction of the middle one.

\begin{figure}[h] 
\centering
\includegraphics[width=0.5\linewidth]{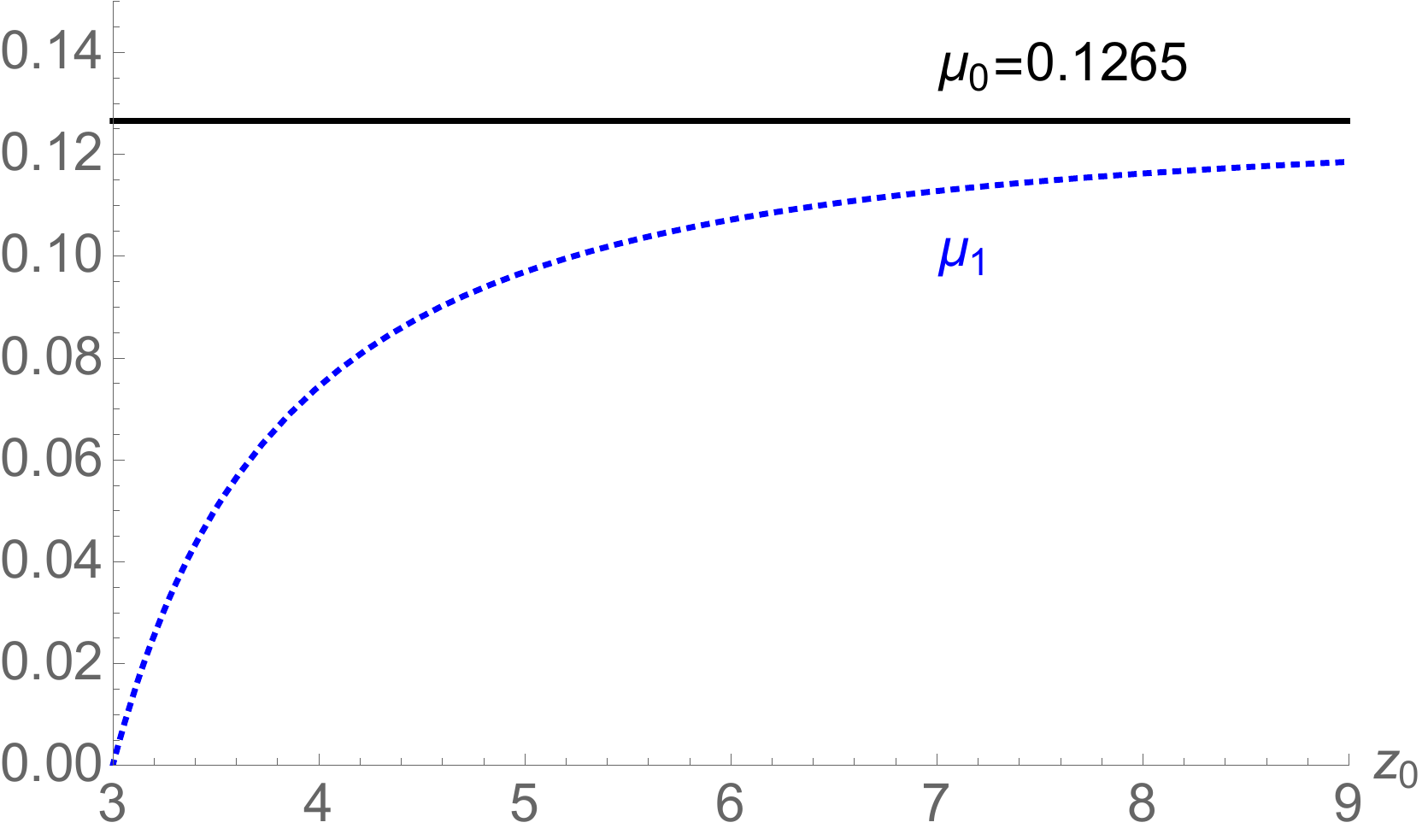}~
\includegraphics[width=0.5\linewidth]{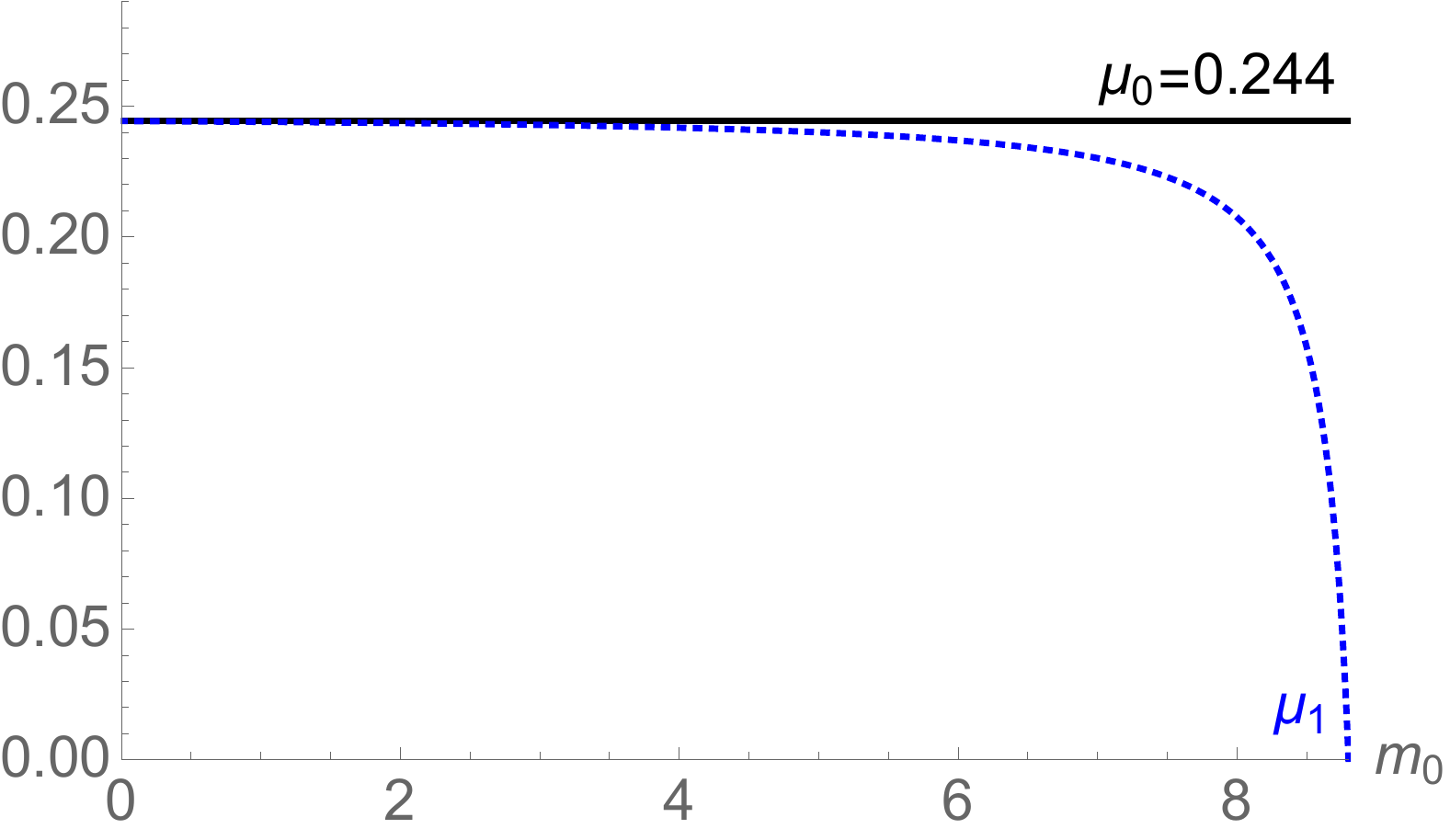}
\includegraphics[width=0.5\linewidth]{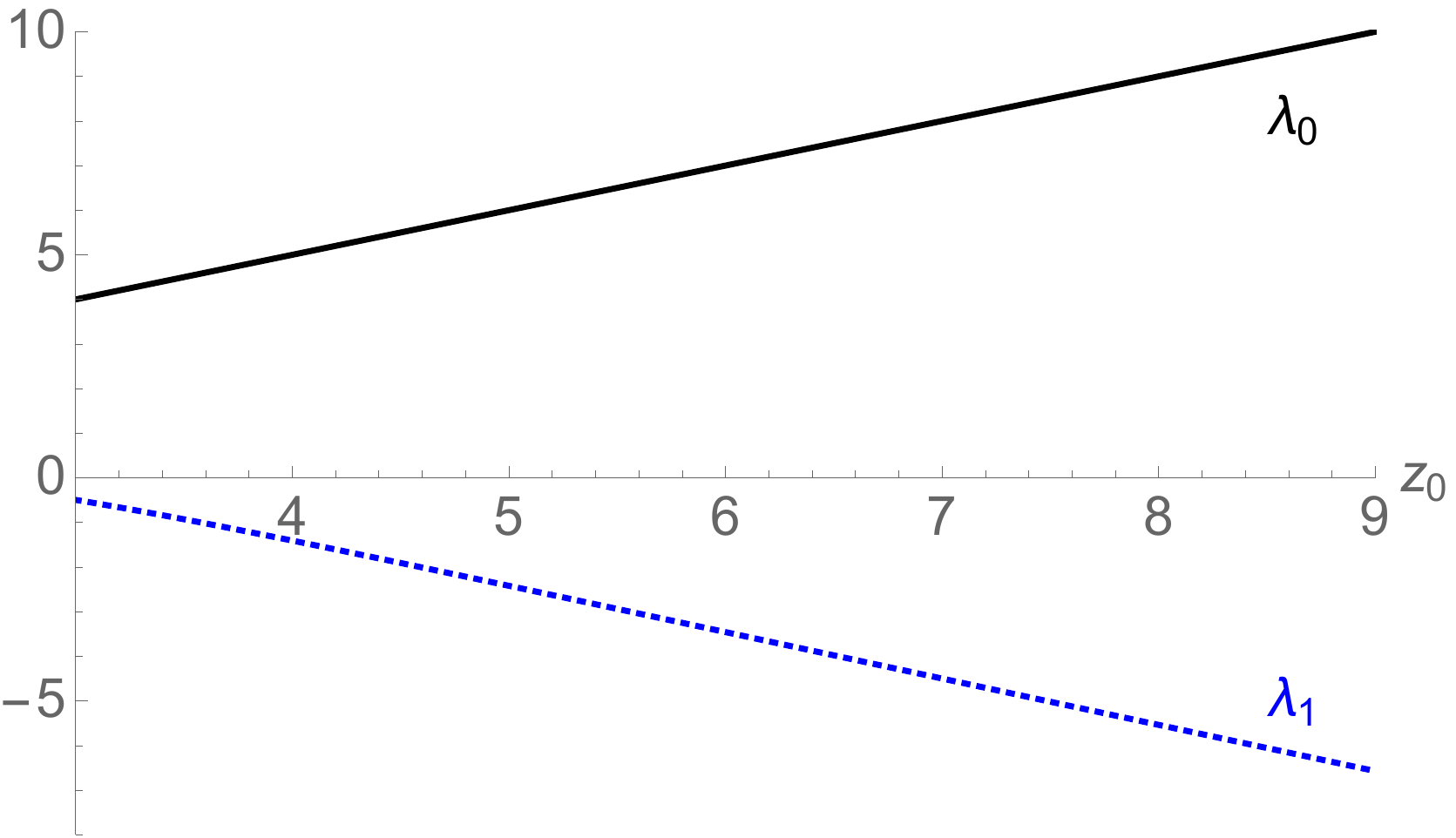}~
\includegraphics[width=0.5\linewidth]{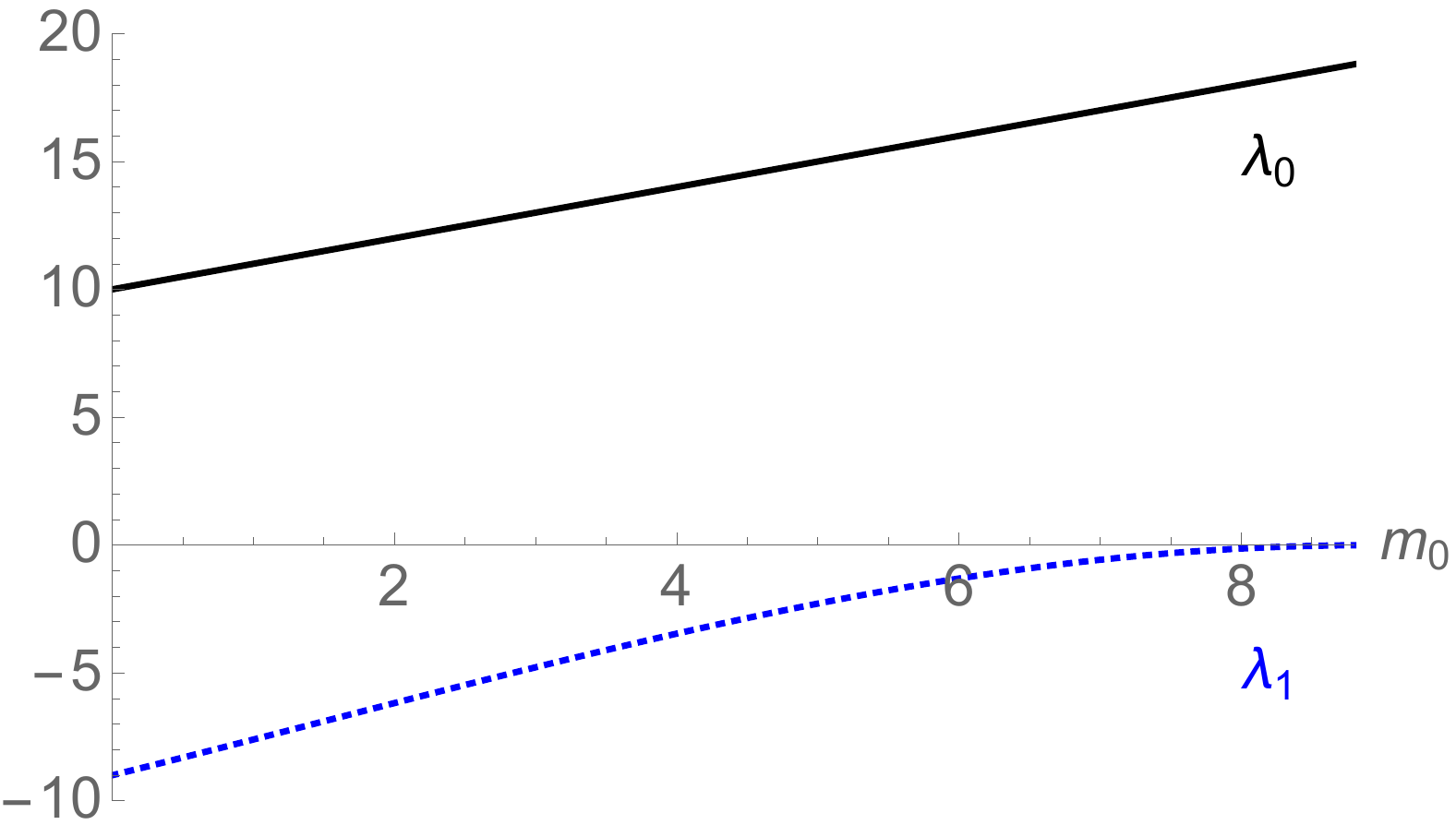}
\caption{The variation of tensions and thermodynamic lengths of the
three black hole system:
Left: for equal masses as a function of black hole separation, and
Right: for fixed black hole separation but varying the mass of the outer
black hole. On the left, the tension is set by taking the minimal value 
consistent with zero tension between the black holes at minimum 
separation, $z_{min}=3$, giving an ambient tension of $41/324$. 
On the right the separation is set at $z_0=10$, and the outer black hole
mass varies from zero to $8.8$, which is very close to the merger limit of 
maximal tension, $\mu_0\sim 0.244$.}
\label{fig:muandlambda}
\end{figure}

For the thermodynamic lengths we have: 
\be
\beal
\lambda_0 &= ( z_6 - z_1)/2 = (z_0+m_0) = \lambda_3 \,,\\
\lambda_1 &= -(z_0-m_0-m)
\frac{(z_0^2-m_0^2)(z_0^2-(m_0+m)^2)}{z_0^2(z_0^2-(m_0-m)^2)} 
=\lambda_2 \,.
\eeal
\ee
Thus the thermodynamic length of the ambient deficit---that is,
the total from both string $1$ and $4$---is
the distance from the north pole of the topmost black hole
to the south pole of the bottom-most black hole.
The length associated to the intermediate strings is 
\textit{minus} the distance between the horizons of adjacent black holes
(see figure \ref{fig:muandlambda}).

We have found an interesting phenomenon where the thermodynamic
lengths of the outer strings are positive
whereas those of interior strings are negative.
This is puzzling from the perspective of the individual black holes.
However, upon taking the system as a composite it makes sense:
if we alter the overall tension, we must account for the contributions 
from both inner and outer cosmic strings.
The negative contribution from the interior lengths then
counteracts the positive contribution from the outer lengths.
Explicitly, first set up the three black holes so that there
is no deficit between the central and outer black holes.
That is, $K$ takes the value
\be
K_0 = \frac{z_0^2(z_0^2-(m_0-m)^2)}{(z_0^2-m_0^2)(z_0^2-(m_0+m)^2)} \,.
\ee
We now ``add'' a cosmic string to the system by increasing $K$ to $K_0+K_1$,
so that
\be
\delta \mu_0 = \frac{K_1}{4K_0(K_0+K_1)} \qquad,\qquad
\delta \mu_1 = \frac{K_1}{4(K_0+K_1)} = K_0 \delta \mu_0 \,.
\ee
Note that the tension of the ambient cosmic string through the whole
spacetime increases from $\mu_0=(1-1/K_0)/4$ to $\mu_0+\delta \mu_0$. 
However, the region between the black holes,
which initially had no deficit,
now exhibits a cosmic string with tension 
$K_0 \delta \mu_0$,
i.e.,\ a slightly greater tension than the increase in ambient string tension. 
Now let us look at the overall change in energy:
\be
\beal
\sum_{I=0}^3 \lambda_I\delta \mu_I &=
\lambda_0\delta \mu_0 + \lambda_1\delta \mu_1 +
\lambda_2\delta \mu_2 + \lambda_3\delta \mu_3 \\
&= 2(z_0+m_0) \frac{K_1}{4K_0(K_0+K_1)}
-2\frac{(z_0-m_0-m)}{K_0} \frac{K_1}{4(K_0+K_1)}\\
&= \frac{(4m_0+2m)K_1}{4K_0(K_0+K_1)} \,.
\eeal
\ee
This is the total length of string captured by the black holes
multiplied by the tension.
We conclude therefore that the thermodynamic lengths 
really do behave in concert,
combining in such a way that the overall modification of tension has a
sensible impact on the overall thermodynamics of the system.

\begin{figure}[h] 
\centering
\includegraphics[width=0.5\linewidth]{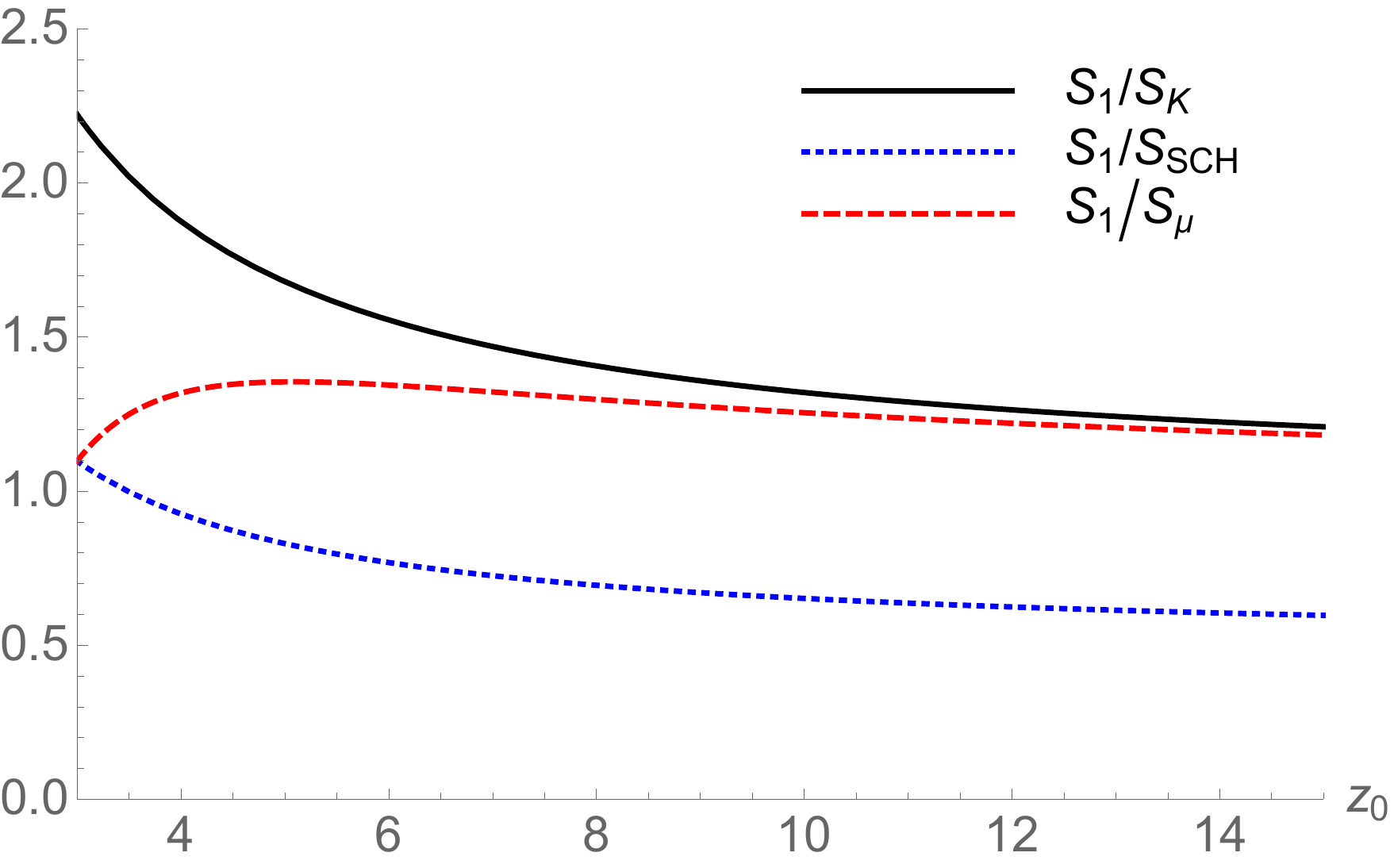}~
\includegraphics[width=0.5\linewidth]{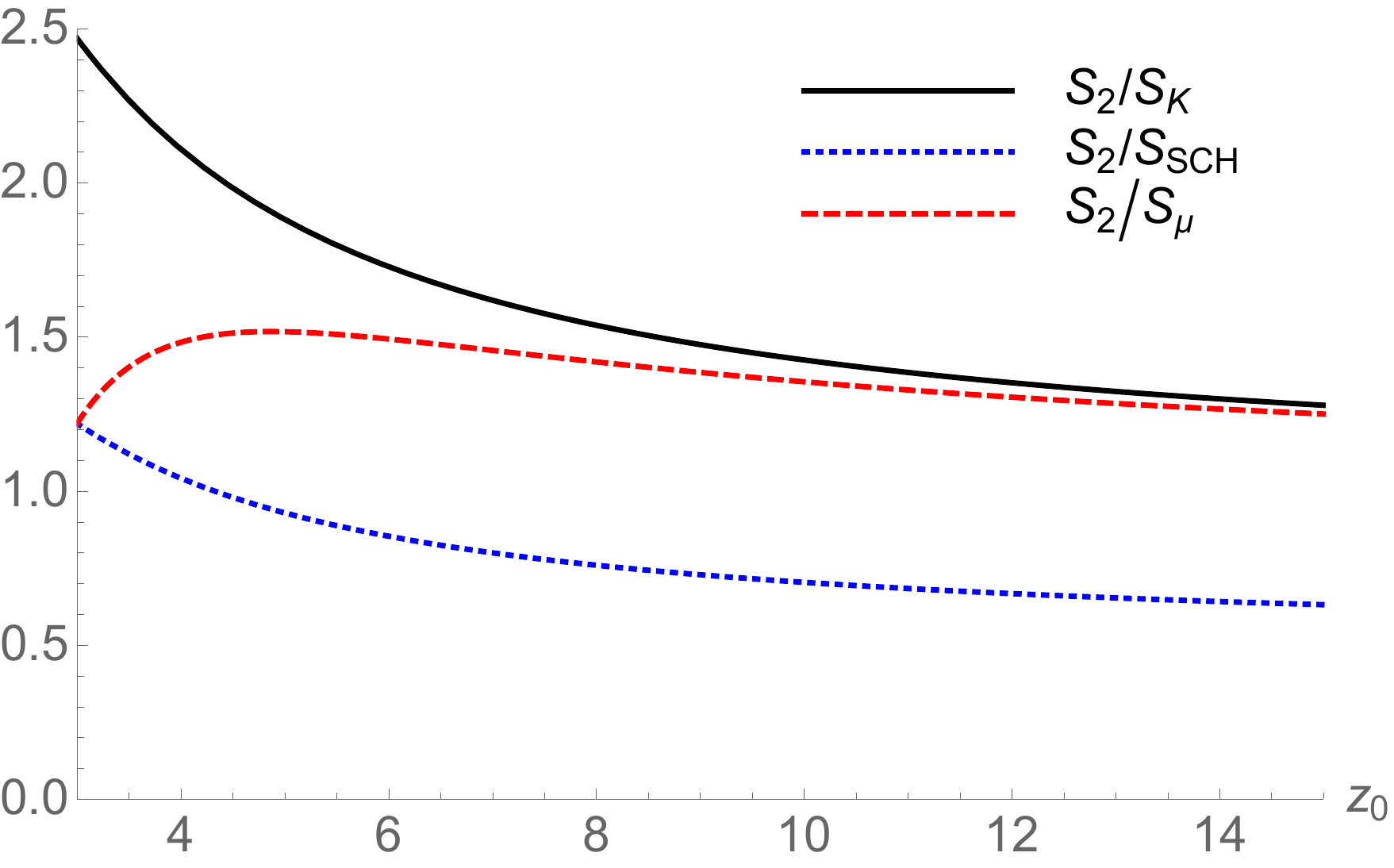}
\caption{The entropy of the outer (Left) and middle (Right) black holes 
for equal masses as a function of black hole separation. The black holes
all have unit mass, with an ambient tension of 41/324. The tension
is set by taking the minimal value consistent with no strut between the
black holes at minimum separation, here $z_{min}=3$. } \label{fig:S2}
\end{figure}
Turning to the entropies, one sees that $S_2/m^2>S_1/m_0^2$.
Essentially, this is saying that the inner black hole has a higher entropy 
in units of its mass (squared) than the outer ones.
We understand this from the impact of the conical deficits:
entropy is decreased in general by having a conical deficit, as part of
the horizon is ``cut out'', leaving a rugby, as opposed to soccer, ball
shape. We would expect that the entropy of the middle black hole
would be relatively higher, as the deficit running through this black hole is less than
the deficit emerging from the outer poles of the outer black holes.

The picture is a little more subtle than this broad brush expectation however;
the central black hole has a uniform tension, $\mu_1$,
running through it, so naively, we might expect that the entropy might be
tracked by $4\pi m^2(1-4\mu_1)$,
but in fact the entropy is higher than this. 
For the outer black holes, we might expect the entropy to be tracked by
the average tension between the poles, but again, it is higher.
Indeed, the entropy is higher even than the Schwarzschild entropy for a 
range of separation values $z_0$; see figure \ref{fig:S2}.

Similarly, we can track what happens to the entropy of one black hole
as a result of changing the mass of the others. For example, keeping
the central black hole at unit mass, and keeping the other two black holes 
at a given distance, we can see how the entropy of the central black hole
\be
S\textsubscript{central}
= \frac{4\pi}{K} \frac{(1-m^2/z_0^2)(1+m/z_0+1/z_0)^2}
{(1-m/z_0+1/z_0)^2}
\label{centralentropy}
\ee
alters as we change the mass of the outer black holes. 
The mass of the outer hole $m_0$
can range from zero to $z_0-1$,
however at this point the horizons
merge and to maintain a non-negative tension between the black holes 
we would have to have a maximal deficit of $2\pi$. Instead, we choose
a maximal mass $m_{\text{max}}$, and set $K$ so that at the maximal 
mass there is no deficit between the black holes:
\be
K = K_c \equiv \frac{z_0^2(z_0^2-(m_{\text{max}}-1)^2)}
{(z_0^2-m_{\text{max}}^2)(z_0^2-(m_{\text{max}}+1)^2)} \,.
\ee

\begin{figure}
\centering
\includegraphics[width=0.48\linewidth]{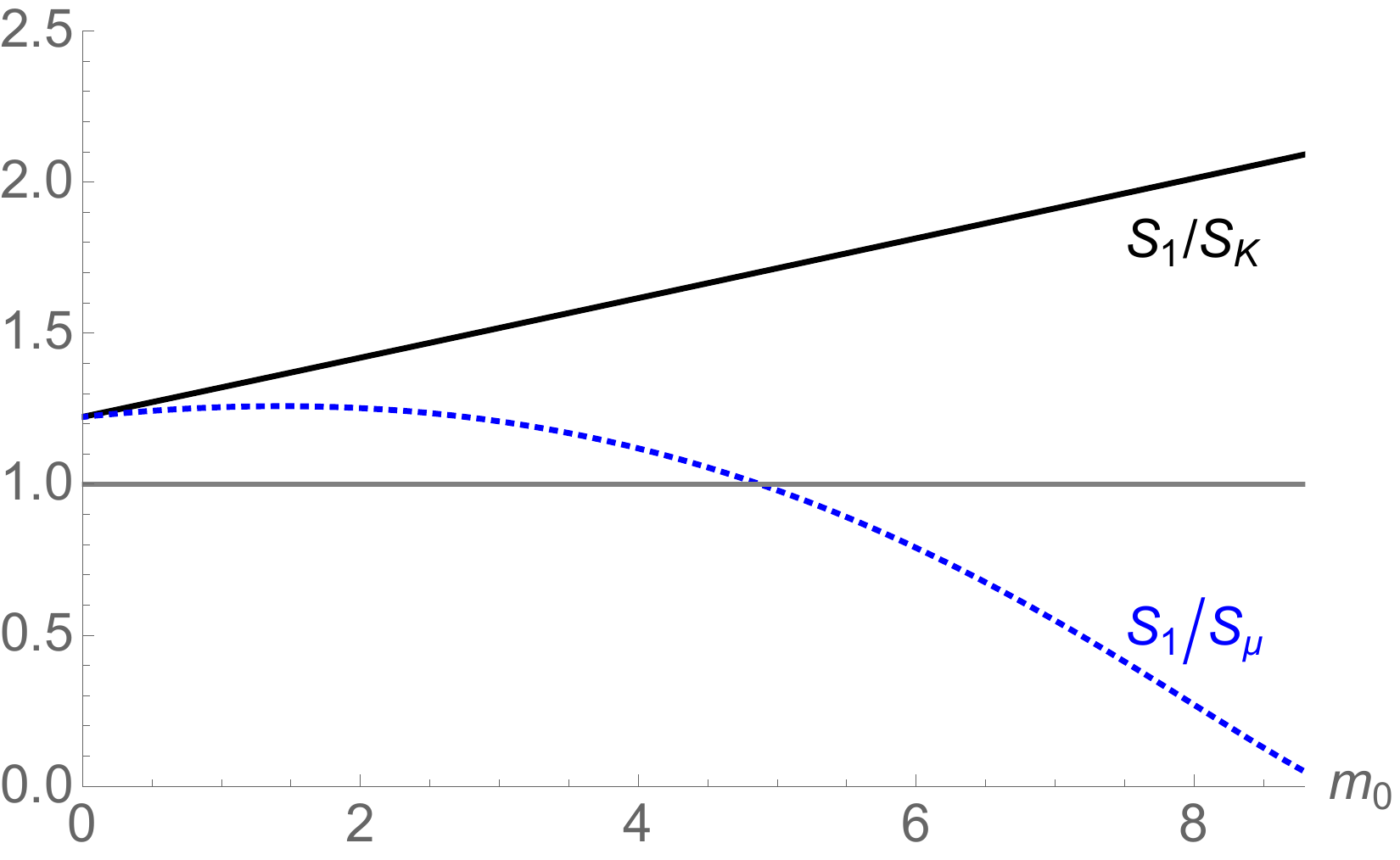}
\includegraphics[width=0.48\linewidth]{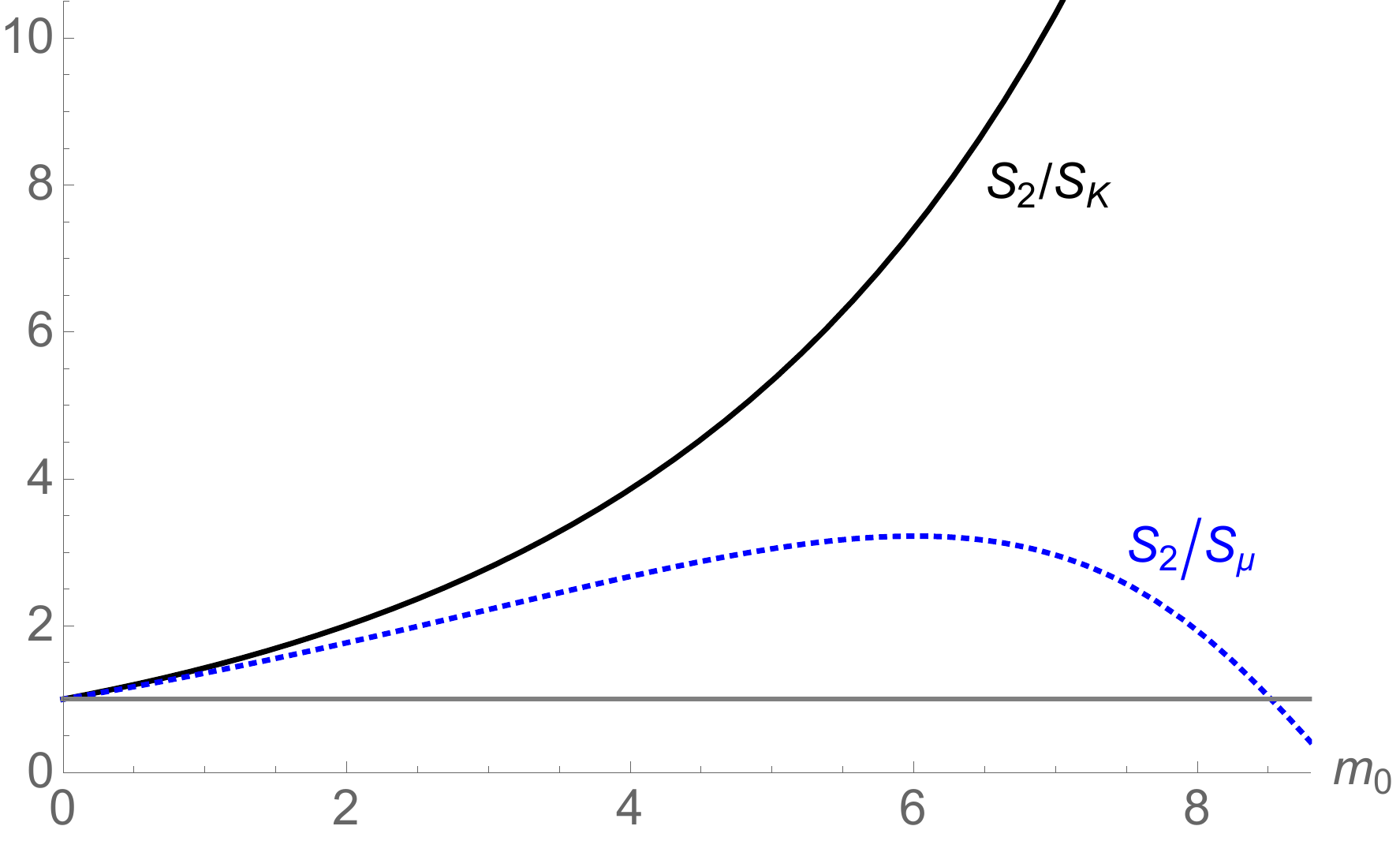}
\caption{The entropies of the outer (Left) and central (Right)
black holes as a function of the mass $m_0$ of the outer black holes.
The mass of the central black hole is fixed at $1$, and the outer black
holes have the same mass $m_0$.}
\label{fig:Smass}
\end{figure}

Figure \ref{fig:Smass} shows the variation of the entropy of the central 
black hole for a separation $z_0 = 10$, and a mass range up to
$m_{\text{max}}=8.8$. This is very close to the merger limit,
giving a large external tension $\mu_0 \sim 0.244$, so a deficit angle of
$\delta/ (2\pi) \sim 0.977$. As before, the entropy is normalised by 
the entropy of a single black hole in a spacetime with both this
ambient deficit ($S_K = 4\pi/K$) as well as that of a black hole
with a cosmic string of tension $\mu_1$ running through 
($S_\mu = 4\pi(1-4\mu_1)$).

We now see a more nuanced behaviour. Initially, at $m=0$, the 
spacetime is precisely that of a single black hole of unit mass pierced
by a cosmic string of tension $\mu_0=\mu_1=(1-1/K)/4$. As we 
switch on the black hole mass at $z_0$, $\mu_1$ decreases,
and this results in an increase in entropy, but this is over and above
what we would expect simply from a drop in $\mu_1$. This
comes primarily from the $m$ dependence in \eqref{centralentropy}.
As we increase the mass further however, while the function $S_2/S_K$
continues to grow, the ratio $S_2/S_\mu$, of the entropy to that of a black
hole with the $\mu_1$ cosmic string starts to drop, eventually becoming
less than one. We can understand this as being a consequence of 
the very large deficit in the majority of the spacetime, even though
locally, at the central black hole, there is no cosmic string. 
The outer black holes are very close (within a Schwarzschild radius)
to the central black hole, thus the geometry is strongly distorted there.

\subsection{Accelerating arrays}

Now let us consider accelerating black hole arrays
of the type depicted in figure \ref{fig:AccSources}.
The main difference with the non-accelerating array is that
we have chosen the parameter $K$ to represent the ambient tension, so
that the asymptotic tension in principle varies
with the locations of the rod ends. 
The expressions for entropy, temperature, tension and thermodynamic length 
are readily worked out from \eqref{eq:entropy} and \eqref{eq:temperature},
though are not particularly illuminating. 
However, we can intuit the general behaviour
as we vary the black hole masses and positions.

First, note that $\mu_0>\mu_N$.
We expect this because since the black holes are accelerating 
there must be an imbalance between the tension of the string
coming in from infinity and that of the string exiting 
through the acceleration horizon.
Next, as we increase the first black hole mass $m_1$, the first tension
$\mu_1$ will drop, as more of the pulling power of the string will be used to
accelerate the increased mass.
Whether the subsequent string tensions increase or decrease 
depends on the masses of the individual black holes:
the second black hole will be attracted to the first (and third, 
if present) which provides an additional attractive force over and above
that of the cosmic string. Typically, if the black holes are well separated
relative to their size, the string tensions will cascade down in magnitude
as one moves along the array, but for large black holes, this need not be
the case (see the two black hole case below).

\subsubsection{The C-metric}

It is worth briefly checking the C-metric results, 
first proposed in \cite{Anabalon:2018qfv}. 
The C-metric has a single horizon and a SILM
so we have $n=3$ and $N=1$.
The metric in Weyl form is
\be
ds^2 = \frac{X_1X_3}{\ell_\gamma X_2} dt^2
- \frac{\ell_\gamma E_{12}E_{23}}{4R_1R_2R_3 E_{13}}
\left( \frac{z_3-z_0}{z_3-z_2} \right)^2 [dr^2+dz^2] 
- r^2 \frac{\ell_\gamma X_2}{X_1X_3}\frac{d\phi^2}{K^2}\,,
\ee
where $z_0 = (z_1+z_2)/2$ is the centre of the black hole rod,
and we have replaced $V_3 = (z_3-z_1)/(z_2-z_2)$.
Here, $\ell_\gamma = 2(z_3 - z_0)$ is shown to be the reciprocal 
of the acceleration of a small black hole in appendix \ref{app:cmet},
where we also note the transformation 
between this metric and the more familiar spherical coordinates.

Turning to the thermodynamics, we compute $z_c$ as
\be
z_c = z_3 - \frac{(z_3-z_2) (z_3-z_1)}{(z_3-z_0)}
= \frac{(z_3z_0-z_1z_2)}{(z_3-z_0)}\,.
\ee
Meanwhile, the entropy and thermodynamic lengths are
\be
\beal
S &= \frac{4 \pi m^2}{K}\frac{(z_3-z_0)^2}{(z_3-z_2)(z_3-z_1)}
&\to& &\frac{4 \pi m^2}{(1-4m^2A^2)}\\
\lambda_0 &= e^{\nu_0}(z_c-z_1)
= m\frac{(z_3-z_1)}{(z_3-z_2)} &\to&& \frac{m(1+2mA)}{(1-2mA)}\\
\lambda_1 &= e^{\nu_1} (z_2-z_c)
= m \frac{(z_3-z_2)}{(z_3-z_1)} &\to& &\frac{m(1-2mA)}{(1+2mA)}
\eeal
\ee
in agreement with the parameters proposed in \cite{Anabalon:2018qfv}.

It is also straightforward to write down a 
Christodoulou-Ruffini-like formula \cite{Christodoulou:1972kt} 
for the C-metric.
Following \cite{RuthAndy}, 
define a quantity $\Delta$ characterising the average tension 
emerging from the black hole horizon,
and a quantity $C$ characterising the tension differential:
\be
\beal
\Delta &= 1-2(\mu_0+\mu_1) = \frac{1}{K}  \,,\\
C & = \frac{\mu_0-\mu_1}{\Delta} = \frac{z_2-z_1}{4\ell_\gamma}
\rightarrow m A    \,.
\eeal
\ee
Then one finds that
\be
M^2 = \frac{\Delta S}{4\pi}\left(1-4C^2\right) \,.
\ee
Increasing the acceleration of the black hole while 
maintaining a constant ambient deficit removes energy from the black hole.
This result is not as unsettling as it may first appear, 
as energy may be lost both across the acceleration horizon
and as gravitational radiation at future infinity \cite{Podolsky:2003gm}.

\subsubsection{Two accelerating black holes}

As a less trivial example, we present results for the two accelerating 
black hole system, first explored in \cite{Fay}. We have 
\be
\beal
\ell_\gamma
&= 2z_5 - \frac{(z_4^2-z_3^2+z_2^2-z_1^2)z_5
- (z_4-z_3+z_2-z_1)(z_1z_3+z_2z_4)}
{(z_4-z_3+z_2-z_1) z_5+z_1z_3-z_2z_4}\\
&\sim 2 (z_5 - z\textsubscript{com}) + \mathcal{O}\left(z_5^{-1}\right)\,,
\eeal
\ee
where
$z\textsubscript{com} = \frac{z_4^2-z_3^2+z_2^2-z_1^2}{2(z_4-z_3+z_2-z_1)}$ 
is the centre of mass of the pair of black holes (this formula generalises to any
number of accelerating black holes).

\begin{figure}[h] 
\centering
\includegraphics[width=0.48\linewidth]{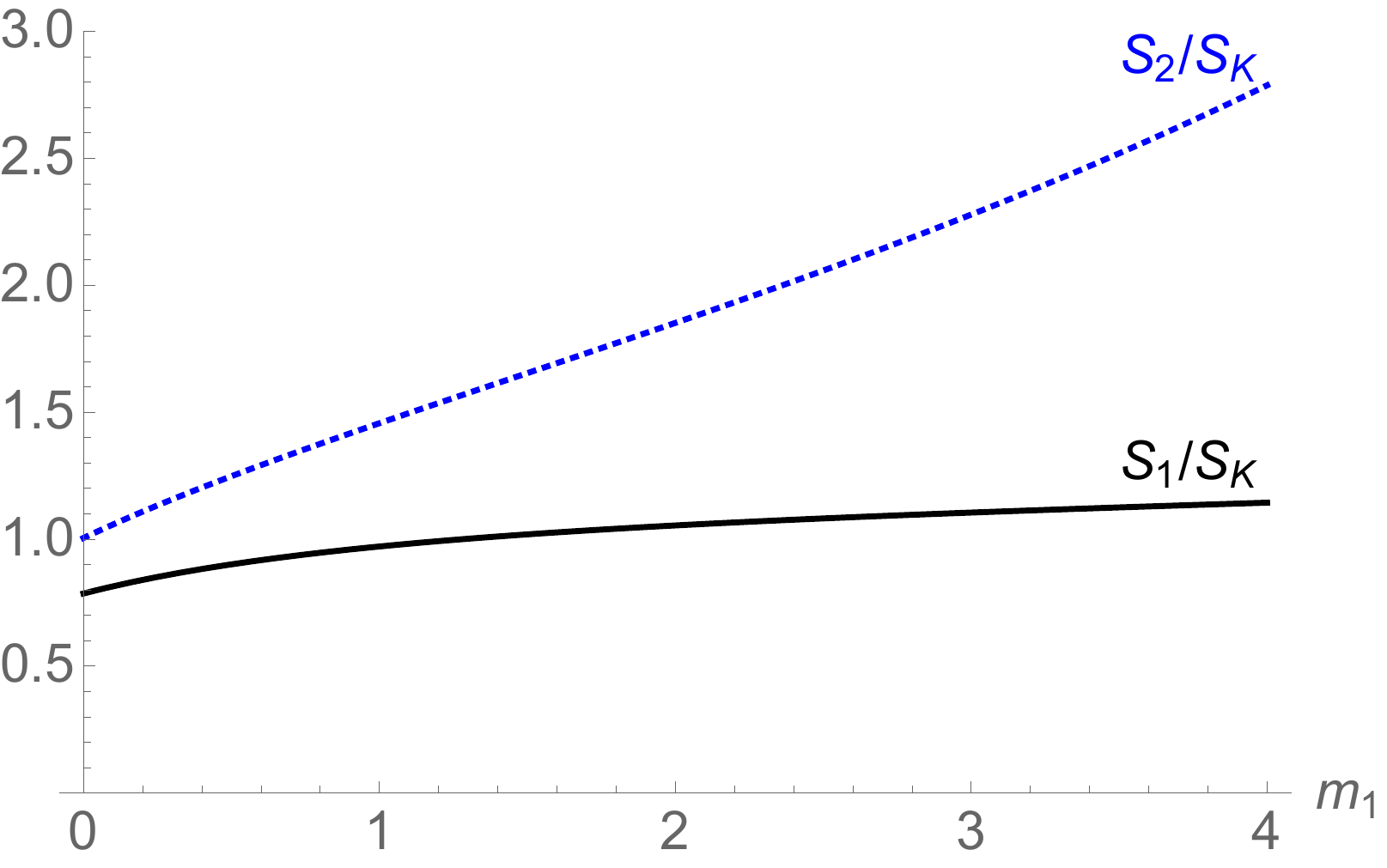}
\includegraphics[width=0.48\linewidth]{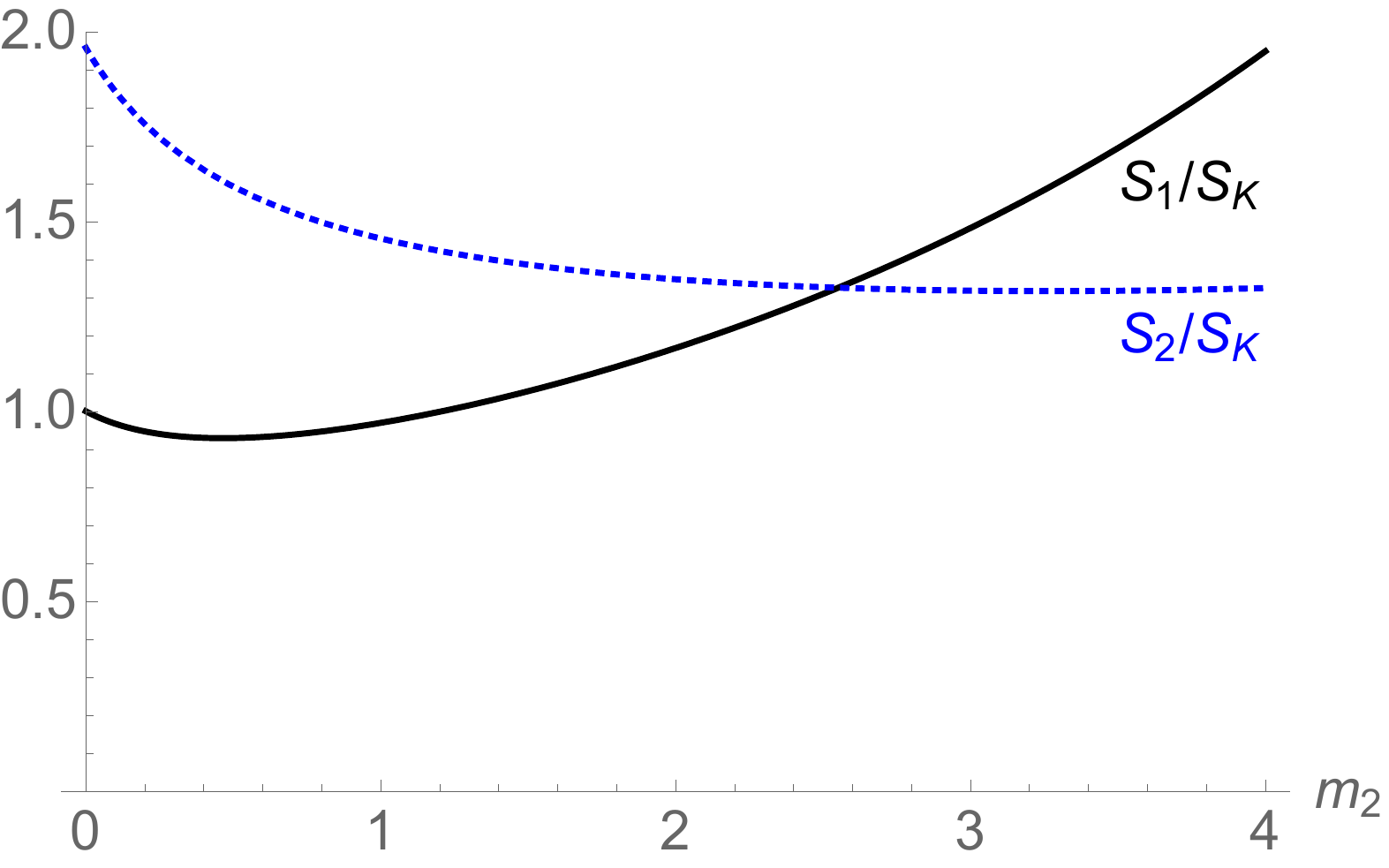}
\includegraphics[width=0.48\linewidth]{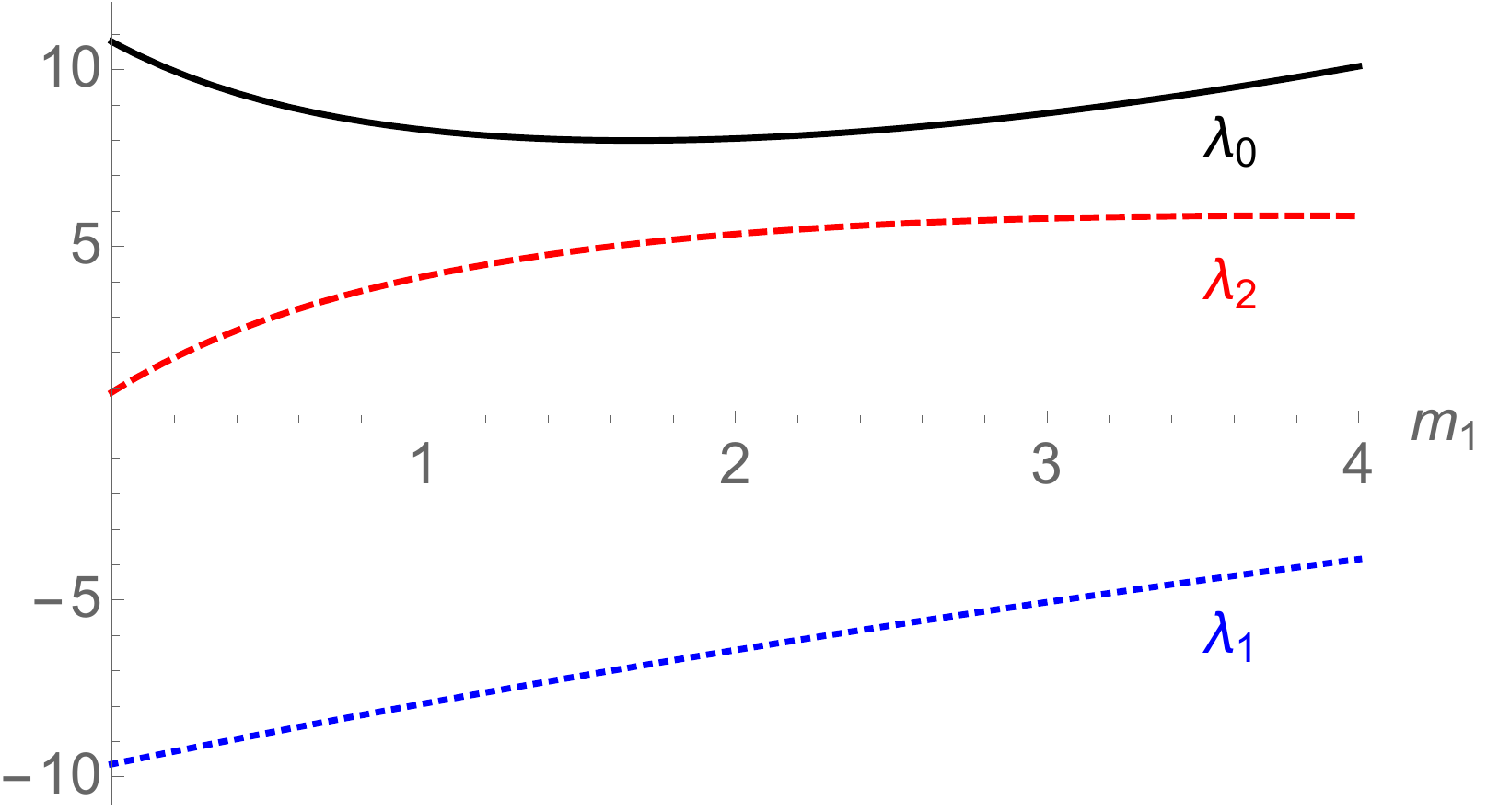}
\includegraphics[width=0.48\linewidth]{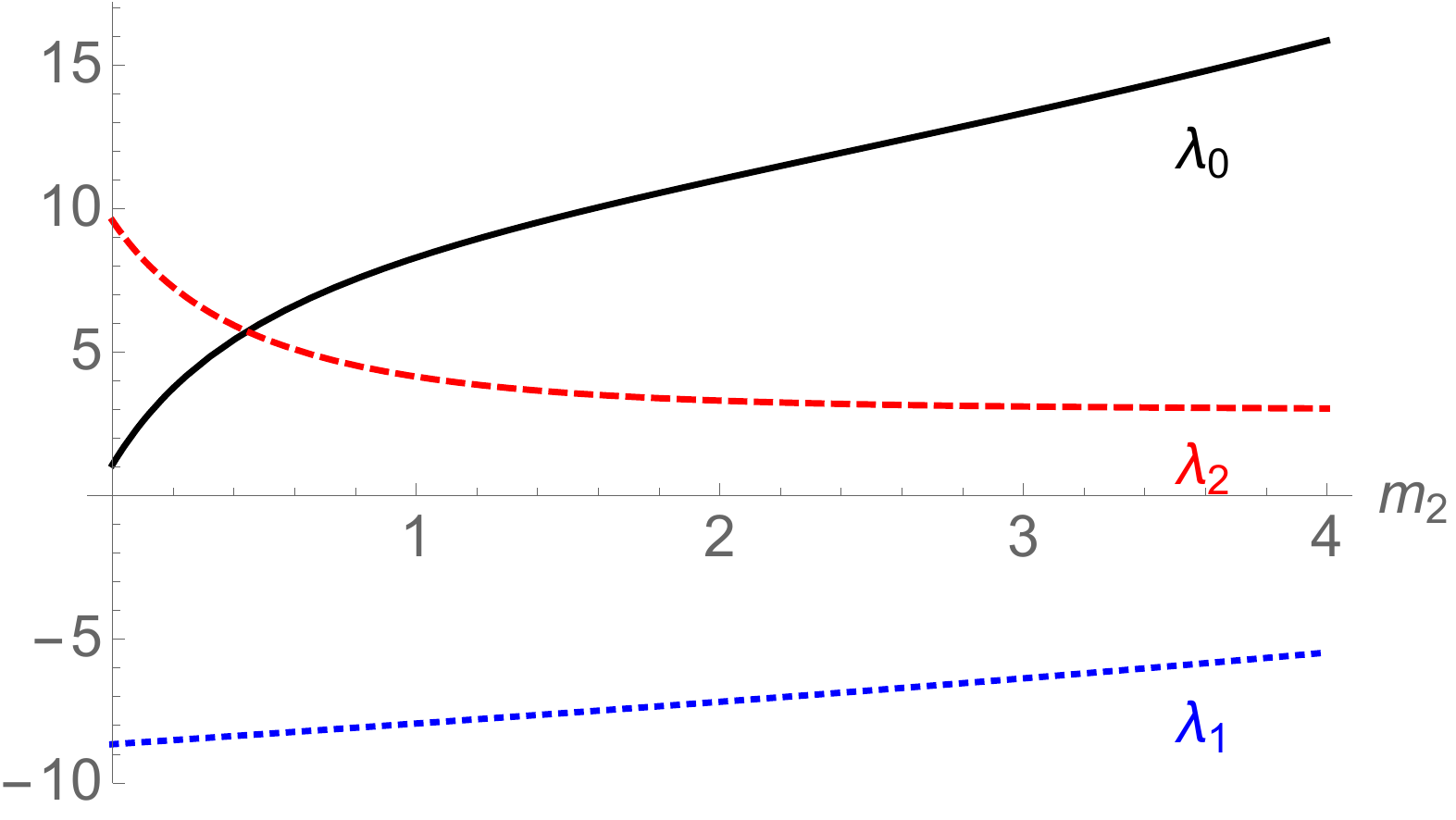}
\caption{The variation of entropies and thermodynamic lengths
as a function of mass in a two accelerating black hole set-up.
The outgoing tension is fixed at $\mu_0=1/8$, and the displacement
of each rod from the origin at $z_b = 5$. One mass is fixed at unity,
with the other mass varying from zero to $4$.
The upper plots show how the entropies, normalized by $4\pi m_I^2/K$,
vary, and the lower plots the thermodynamic length. 
Note that $K$ varies as $m_I$ varies, in order to keep $\mu_0$ fixed.} 
\label{fig:twoaccbhvarym}
\end{figure}

Placing the two black holes at $\pm z_b$ fixes the gauge, and we can 
see how the string tensions and black hole entropies react to changes in
black hole mass and distance to the horizon (without loss of generality
we can keep $z_b$ fixed as a choice of scale). Writing 
\be
S_0 = \frac{(m_1+m_2)(m_1+m_2+2z_b)(m_1m_2-z_b^2+z_5^2)^2}
{(z_5+z_b+m_1)(z_5-z_b-m_2)(z_5(m_2+m_1) + z_b(m_2-m_1))}  \,,
\ee
the entropies are
\be
\beal
S_1 = \frac{4\pi m_1^2}{K}
\frac{S_0}{(2z_b+m_1-m_2)(z_5+z_b-m_1)} \,, \\
S_2 = \frac{4\pi m_2^2}{K}
\frac{S_0}{(2z_b+m_2-m_1)(z_5-z_b+m_2)} \,.
\eeal
\label{s12acc}
\ee
We can quickly see that if $m_1=m_2$, the entropy of the first black hole
will always be less than that of the second, which would be expected as
the mean deficit through the first black hole is greater than that through the 
second. However, normalising the entropies with respect to their 
reference $S_K=4\pi m_I^2/K$, we can see that the multiplicative factors 
in \eqref{s12acc} show that both initially decrease as $m_2$ increases 
from zero before turning, although $S_2/S_K$ shows a sharper decrease
and eventually drops below $S_1/S_K$. Again, this behaviour is easy to
see from the ratios in \eqref{s12acc}. 
Figure \ref{fig:twoaccbhvarym} shows this behaviour with varying $m_I$.

In order to compare the impact of varying the masses of the black holes
and their separation, we first fix the outgoing tension at $z\to-\infty$, so
that we are comparing the same conical asymptotics. 
Figure \ref{fig:twoaccbhvarym} shows the effect of
varying the mass of the inner and outer black hole respectively
on the entropies and thermodynamic lengths.
In each case, we fix one of the masses at unity and vary the other.
In both cases, varying the mass of the black hole closer to the 
acceleration horizon ($m_2$) causes a ``crossover'' behaviour.

\begin{figure}[h] 
\centering
\includegraphics[width=0.33\linewidth]{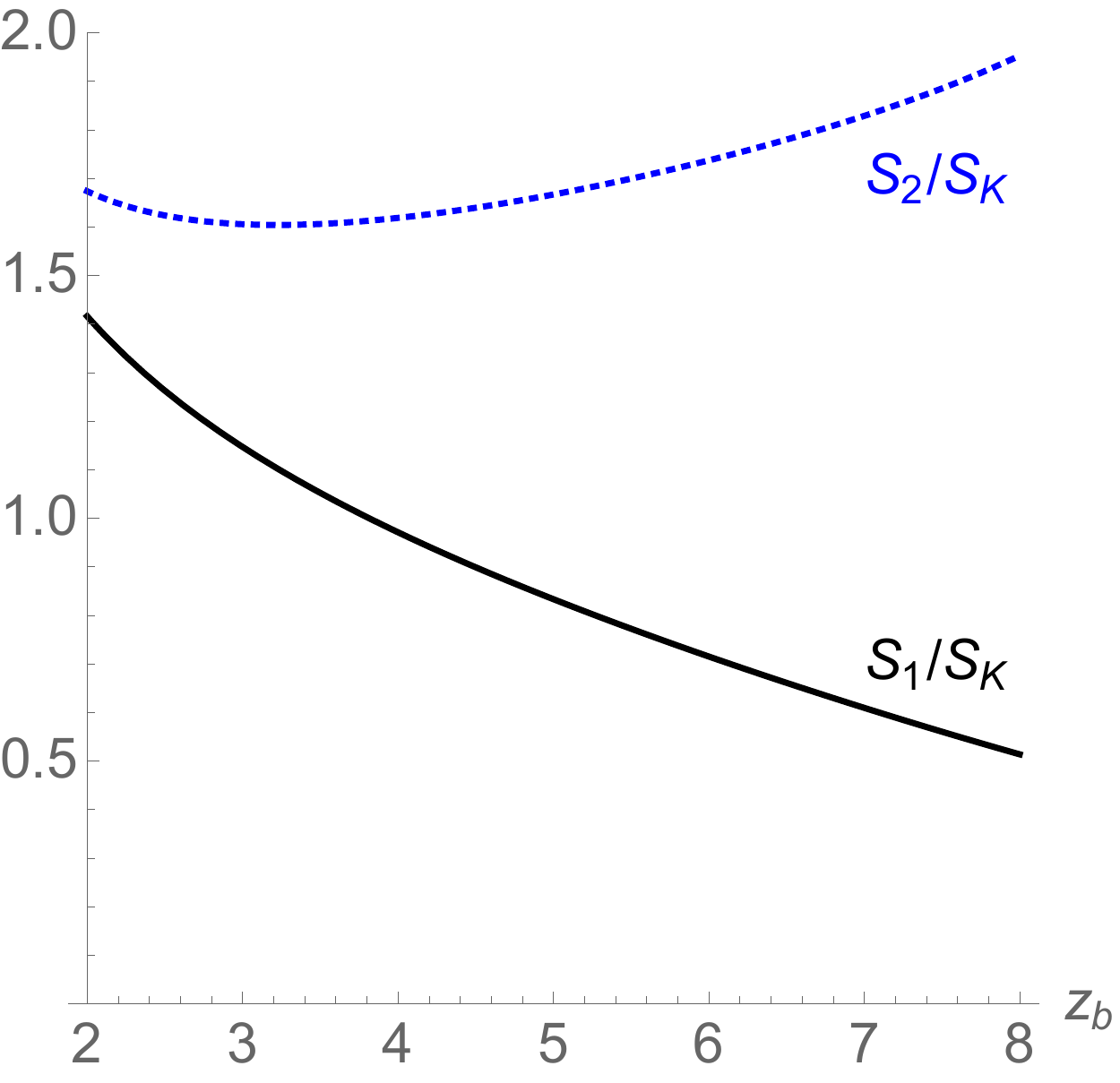}~
\includegraphics[width=0.33\linewidth]{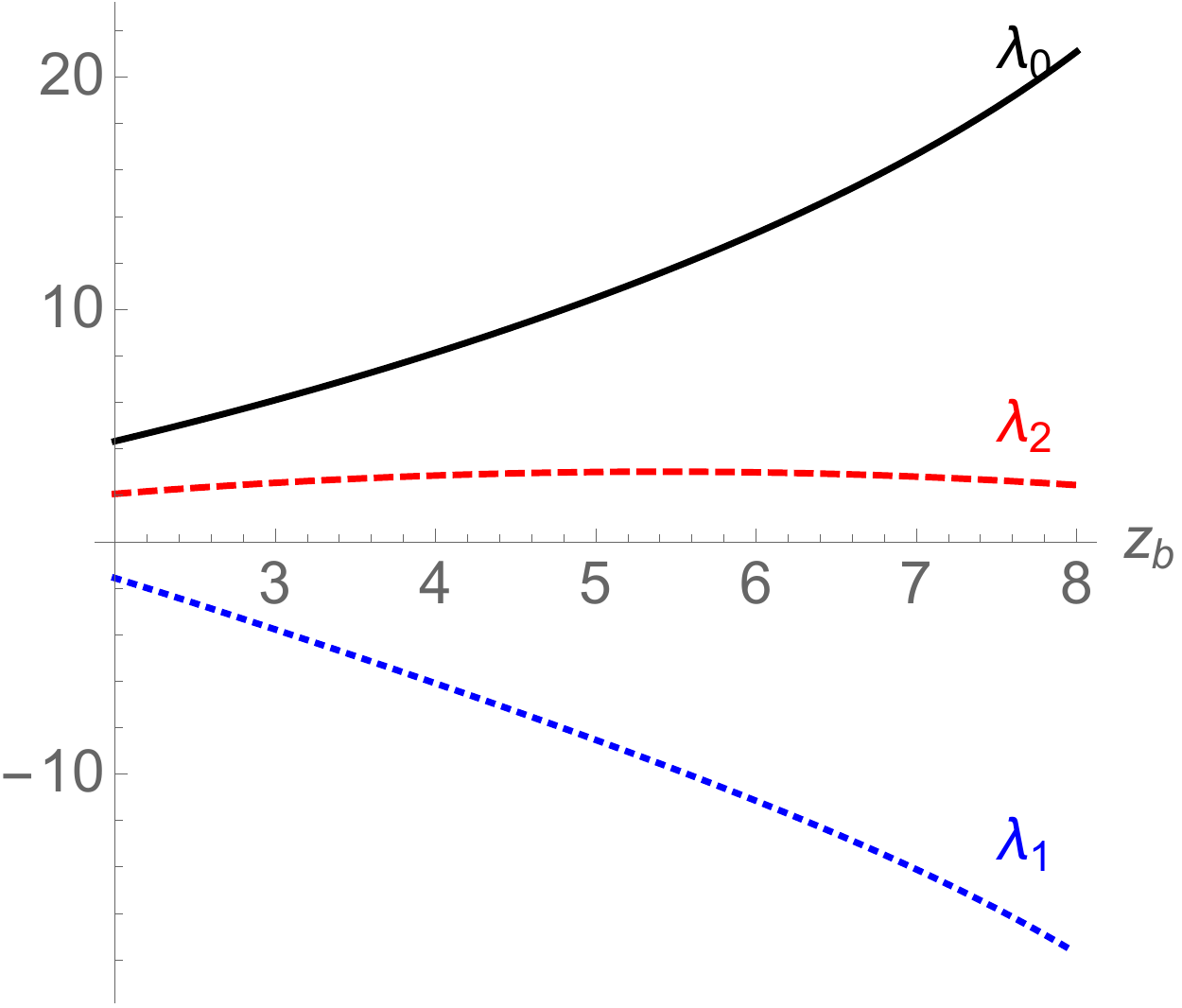}~
\includegraphics[width=0.33\linewidth]{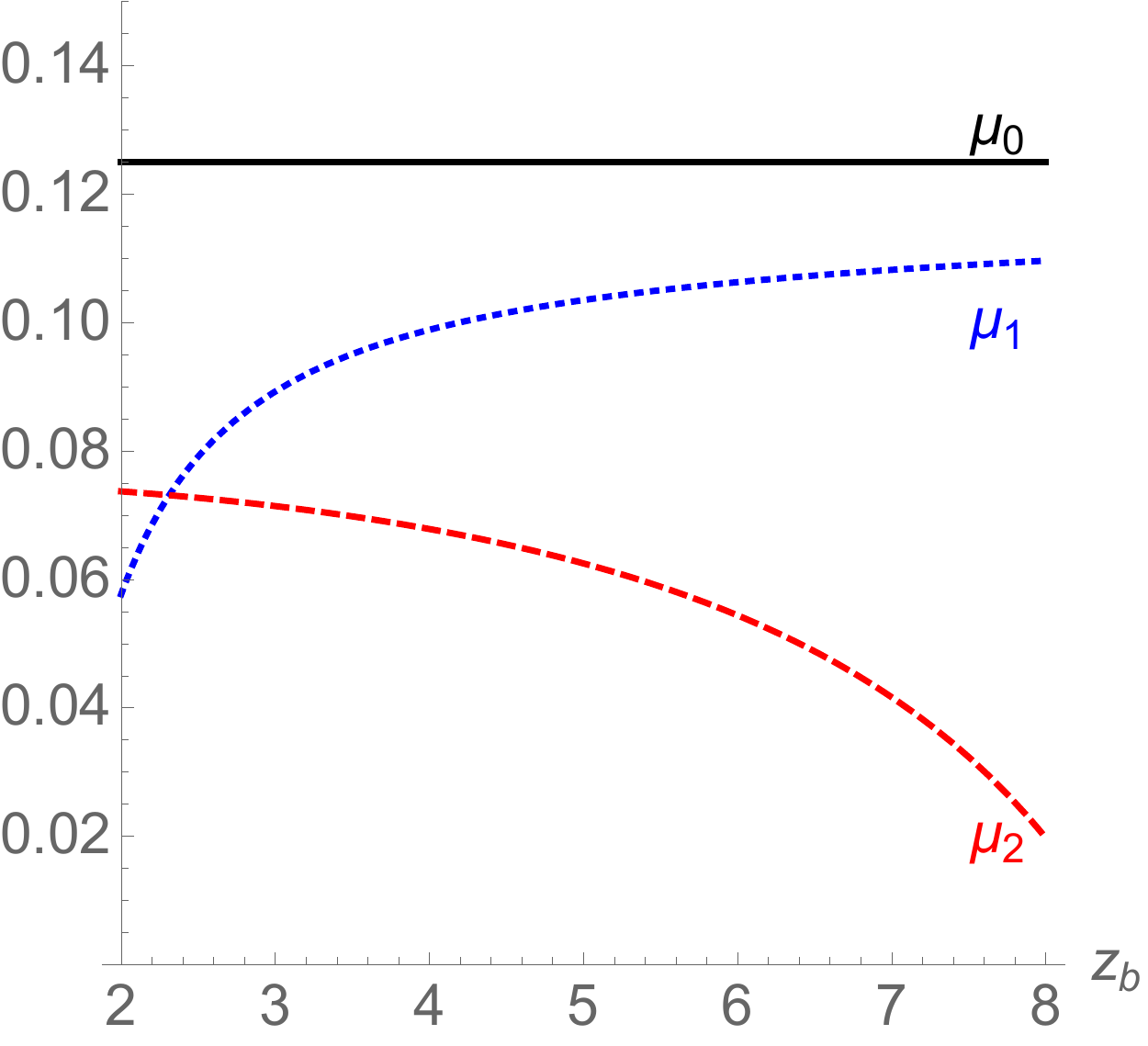}
\caption{The variation of thermodynamic parameters for the double
accelerating black hole set-up where the two black holes have equal mass
$m_1=m_2=1$ and the distance between them $2z_b$ is varied. 
The acceleration horizon is fixed at $z_5=12$.} 
\label{fig:twoaccbhvaryzb}
\end{figure}

Figure \ref{fig:twoaccbhvaryzb} shows how the entropy, length (and
tension) are affected by moving the black holes apart. As before,
the outgoing tension is fixed at $1/8$, and both black hole masses
are fixed at $m_1=m_2=1$; the acceleration horizon is at $z_5=12$.
The normalised entropy of the black hole closer to the acceleration 
horizon increases as the black holes are moved apart, whereas
the entropy of the other black hole decreases sharply. We can get
a rough understanding of this by looking at the string tensions; 
the tension between the second black hole and the acceleration 
horizon drops off sharply at large separation,
meaning that less of the angular direction is cut out by the deficit, thus
increasing entropy. The tension between the black holes, $\mu_1$, in
contrast increases, leading to an expectation that the first entropy
will decrease. While these statements are broadly true -- note that we have
already normalised the $K$ factor out of the entropy, indicating that the
effect of this geometry is magnified.
As expected, the thermodynamic lengths exhibit a scaling with 
increasing separation, with the intermediate length $\lambda_1$
negative and decreasing to compensate the increase in $\lambda_0$.

\section{Conclusions}

To sum up: we have proven a thermodynamic First Law for a 
composite system of black holes, both accelerating and isolated.
We have allowed the varying of the tensions of the cosmic
strings along the axis that are necessary for maintaining 
the equilibrium configuration. As with the accelerating AdS
black hole thermodynamics previously developed, these
strings have a corresponding potential, the thermodynamic length,
which has a direct specification in terms of the Weyl coordinate 
parametrising the axis of symmetry of the black hole array.

We have presented a range of accelerating and non-accelerating
black hole systems
to illustrate the various facets of the thermodynamic parameters.
The main point is that the black holes form a fully composite thermodynamical
system---the variation of one black hole affects all the others. We also see
how the tensions and lengths in a composite system collude in such a way that
the overall picture makes intuitive sense, whereas the individual black hole
contributions may be less transparent. 

Our findings, that the thermodynamic lengths between compact horizons
is related to the proper distance along the axis,
are in agreement with previous results 
\cite{Krtous:2019,Ramirez-Valdez:2020,Garcia-Compean:2020}.
However, in our construction there are also semi-infinite strings 
for which this proper distance would be infinite, yet this is not what 
we would expect thermodynamically. The thermodynamic length
represents the contribution to the enthalpy from the tension (negative
pressure) of the cosmic string inside the black hole, thus should be
finite. We take this into account via a renormalisation process, the $z_c$,
similar to the renormalising of the metric coefficients.

In fact, the result for $\lambda$ derived in \cite{Krtous:2019},
can actually be understood in terms of the covariant-phase space formalism 
\cite{Harlow:2019},
much as black hole entropy and temperature were interpreted
by Iyer and Wald
\cite{Wald,IyerWald1}.
In this construction, the idea is that, on shell, variations
the action consist solely of boundary data.
Taking this variation to be the action of some Killing vector field
corresponding to time translation,
one can find a quantity which vanishes when integrated over a Cauchy slice.
Taking the variation of this quantity, and splitting the integral up into 
boundary pieces via Gauss' law, one obtains the First Law.
The contribution from infinity gives the variation in mass
and the contribution at the horizon gives $T\delta S$.
When strings are present, one must also consider the contribution
from a new surface: a ``tube'' which encases the string\footnote{Similar
ideas have been applied to thermodynamic investigations of
black holes possessing Misner strings \cite{NUT1,NUT2}.}.
It is precisely this contribution which provides $\lambda\delta\mu$.
From this perspective, the thermodynamic lengths calculated in 
\cite{Anabalon:2018ydc,Anabalon:2018qfv}
for the AdS C-metric may be seen as renormalised worldvolumes
(per unit time) of the infinite proper length strings.
The sense in which the external strings are renormalised in the asymptotically
flat case is less clear and would be interesting to understand.

While the system of many black holes is not stable, 
it is nonetheless interesting that it too 
displays sensible thermodynamic properties, 
further supporting 
the inclusion of cosmic strings in the thermodynamic picture.

\acknowledgments

This work was supported in part by the
STFC [Consolidated Grant ST/P000371/1 -RG, DTG - AS], 
and by the Perimeter Institute for Theoretical Physics (RG). 
Research at Perimeter Institute is supported by 
the Government of Canada through the Department of Innovation, 
Science and Economic Development Canada and by the Province of 
Ontario through the Ministry of Research, Innovation and Science.

\appendix
\section{Coordinate systems for the C-metric}\label{app:cmet}

We collect the transformation formulae between the standard C-metric
(expressed in Hong-Teo form \cite{Hong:2003gx})
and the Weyl form of section \ref{sec:weyl}.
The C-metric in Weyl form is
\be
ds^2 = \frac{X_1X_3}{\ell_\gamma X_2 } dt^2
- \frac{\ell_\gamma E_{12}E_{23}}{4R_1R_2R_3 E_{13}}
\left( \frac{V_3+1}{2} \right)^2 [dr^2+dz^2] 
- r^2 \frac{\ell_\gamma X_2}{X_1X_3}\frac{d\phi^2}{K^2} \,,
\label{CWeyl}
\ee
where $\ell_\gamma = z_3 - z_0 = z_3- (z_1+z_2)/2$ is the $z$-distance
to the centre of the black hole rod.

Now let $m = (z_2-z_1)/2$ and $A = 1/\ell_\gamma$. Define
\be
\beal
r = {\bar{r}} \sin\theta \frac{\sqrt{f( {\bar{r}} )g(\theta)}}
{(1+A {\bar{r}} \cos\theta)^2}  \,,\quad
z-z_0= {\bar{r}} \frac{(A {\bar{r}} 
+ \cos\theta)(1-m/ {\bar{r}} +mA\cos\theta)}
{(1+A {\bar{r}} \cos\theta)^2} \,,
\eeal
\ee
where
\be
f(R) = (1-A^2 {\bar{r}} ^2)\left(1- \frac{2m}{ {\bar{r}} }\right) \,, \qquad
g(\theta) = (1+2mA\cos\theta) \,.
\ee
Then the Weyl metric \eqref{CWeyl} transforms to the C-metric
in Hong-Teo coords \cite{Hong:2003gx},
rather than the standard Kinnersley-Walker 
coordinates discussed in \cite{Fay}:
\be
ds^2 = \frac{1}{(1+A\bar{r} \cos\theta)^2} \left [ \bar{f}(\bar{r}) dt^2 
- \frac{d\bar{r}^2}{\bar{f}(\bar{r})} 
- \bar{r}^2 \left ( \frac{d\theta^2}{\bar{g}(\theta)} 
- \bar{g}(\theta) \sin^2 \theta d\phi^2\right)\right] \,.
\ee
Here we see the direct interpretation of $\ell_\gamma$
as the acceleration length scale;
for small $m$, $A=1/\ell_{\gamma}$ corresponds to the magnitude of the 
four-acceleration of the black hole \cite{PavelInterpreting}.

\end{document}